\documentclass[a4paper,11pt]{article}
\pdfoutput=1 % to allow pdflatex compilation in JCAP
\usepackage{jcappub}
\usepackage{amsmath}
\usepackage{graphicx}
\usepackage{latexsym}
\usepackage{xspace}
\usepackage{color}
\usepackage{hyperref} 
\usepackage{bm}
\usepackage{relsize}
\usepackage{tabularx}
\usepackage{multirow}
\usepackage{amssymb}
\usepackage[table]{xcolor}
\usepackage{braket}
\usepackage{comment}
\usepackage[utf8]{inputenc}
\usepackage{slashed}
\makeatletter
\gdef\@fpheader{}
\g@addto@macro\bfseries{\boldmath}
\makeatother

\newcommand{\prn}[1]{\left( {#1} \right)}
\newcommand{\com}[1]{\left[ {#1} \right]}
\newcommand{\vev}[1]{ \left< {#1} \right> }

\newcommand{\calN}{{\mathcal N}}

\newcommand{\uw}{\mathrm{w}}
\newcommand{\ubw}{\mathrm{bw}}

\newcommand{\uwell}{\mathrm{well}}

\newcommand{\utot}{\mathrm{tot}}

\newcommand{\Pfpt}{P_{\sss{\mathrm{FPT}}}}

\newcommand{\ie}{\textsl{i.e.~}}

\newcommand{\eg}{\textsl{e.g.~}}

\DeclareMathOperator{\erf}{erf}

\newcommand{\dd}{\mathrm{d}}
\newcommand{\ee}{e}

\newcommand{\sss}[1]{{\scriptscriptstyle{#1}}}
\newcommand{\boldmathsymbol}[1]{{\ensuremath{\boldsymbol{#1}}}}

\newcommand{\uPl}{\mathrm{Pl}}
\newcommand{\uin}{\mathrm{in}}

\newcommand{\uend}{\mathrm{end}}

\newcommand{\ucl}{\mathrm{cl}}

\newcommand{\uc}{\mathrm{c}}

\newcommand{\uS}{\mathrm{S}}

\newcommand{\usssS}{\sss{\uS}}

\newcommand{\usssPl}{\sss{\uPl}}

\newcommand{\nS}{n_\usssS}

\newcommand{\calP}{\mathcal{P}}

\newcommand{\Rea}{\Re \mathrm{e}\,}

\newcommand{\GeV}{\mathrm{GeV}}

\newcommand{\Mpc}{\mathrm{Mpc}}

\newcommand{\Mp}{M_\usssPl}

\newcommand{\efolds}{$e$-folds}
\newcommand{\efold}{$e$-fold}

\newcommand{\beq}{\begin{equation}}
\newcommand{\eeq}{\end{equation}}
\newcommand{\bea}{\begin{eqnarray}}
\newcommand{\eea}{\end{eqnarray}}
\newlength{\wsingfig}
\newlength{\wdblefig}
\newlength{\wquadfig}
\newlength{\wtriplefig}
\newcommand{\Eq}[1]{Eq.~(\ref{#1})}
\newcommand{\Eqs}[1]{Eqs.~(\ref{#1})}
\newcommand{\Fig}[1]{Fig.~{\ref{#1}}}

\newcommand{\Refa}[1]{Ref.~{\cite{#1}}}
\newcommand{\Refs}[1]{Refs.~{\cite{#1}}}
\newcommand{\Sec}[1]{Sec.~\ref{#1}}
\newcommand{\Secs}[1]{Secs.~\ref{#1}}
\newcommand{\App}[1]{Appendix~\ref{#1}}

\subheader{}

\title{Power spectrum in stochastic inflation}

\author[a]{Kenta Ando,}
\affiliation[a]{ICRR, University of Tokyo, Kashiwa, 277-8582, Japan}
\emailAdd{ando@icrr.u-tokyo.ac.jp}

\author[b]{Vincent Vennin}
\affiliation[b]{Laboratoire Astroparticule et Cosmologie, CNRS, Universit\'e de Paris, 75013 Paris, France}
\emailAdd{vincent.vennin@apc.in2p3.fr}

\date{today}

\begin{document}
\sloppy

\abstract{
We compute the power spectrum of curvature perturbations in stochastic inflation. This combines the distribution of first crossing times through the end-of-inflation surface, which has been previously studied, with the distribution of the fields value at the time when a given scale crosses out the Hubble radius during inflation, which we show how to compute. This allows the stochastic-$\delta N$ formalism to make concrete contact with observations. As an application, we study how quantum diffusion at small scales (arising e.g. in models leading to primordial black holes) affects the large-scale perturbations observed in the cosmic microwave background. We find that even if those sets of scales are well separated, large effects can arise from the distortion of the classical relationship between field values and wavenumbers brought about by quantum diffusion near the end of inflation. This shows that cosmic microwave background measurements can set explicit constraints on the entire inflationary potential down to the end of inflation.
}

%\keywords{quantum field theory on curved space, physics of the early universe}

\arxivnumber{2012.02031}

%%%
\maketitle
%%%

%\tableofcontents

%\flushbottom

%%%%%%%%%%%%%%%%%%%%%%%%%%%%%%%%%%%%%%%%%%%%%%%%%%%%%
%%%%%%%%%%%%%%%%%%%%%%%%%%%%%%%%%%%%%%%%%%%%%%%%%%%%%
\section{Introduction}
\label{sec:intro}
%%%%%%%%%%%%%%%%%%%%%%%%%%%%%%%%%%%%%%%%%%%%%%%%%%%%%
%%%%%%%%%%%%%%%%%%%%%%%%%%%%%%%%%%%%%%%%%%%%%%%%%%%%%

Stochastic inflation~\cite{Starobinsky:1982ee, Starobinsky:1986fx} is a powerful formalism to investigate quantum effects in inflating space times. It consists of an effective theory for the long-wavelengths part of quantum fields, when coarse-grained at super-Hubble scales. Above the Hubble radius indeed, the decaying modes become negligible, and possible effects from the non-vanishing commutators between the fields and their conjugate momenta suffer the same fate~\cite{Polarski:1995jg, Lesgourgues:1996jc, Kiefer:2008ku, Martin:2015qta}. Quantum fields thus behave in a classical, though stochastic way (even if they may retain some genuine quantum properties~\cite{Campo:2005sv, Martin:2015qta, Maldacena:2015bha, Martin:2017zxs,  Ando:2020kdz, Green:2020whw}), and evolve according to stochastic Langevin equations. 

The stochastic formalism has been shown to provide excellent agreement with usual quantum field theoretic (QFT) techniques in regimes where the two methods can be compared~\cite{Starobinsky:1994bd, Tsamis:2005hd, Finelli:2008zg, Garbrecht:2013coa, Vennin:2015hra, Onemli:2015pma, Burgess:2015ajz, Vennin:2016wnk, Hardwick:2017fjo, Tokuda:2017fdh, Kitamoto:2018dek, Markkanen:2019kpv}. However the stochastic approach can go beyond perturbative QFT and describe the non-perturbative evolution of the coarse-grained fields through the full nonlinear equations of general relativity. 
It then relies on the separate universe approach~\cite{Lifshitz:1960, Starobinsky:1982mr, Comer:1994np, Wands:2000dp, Khalatnikov:2002kn, Lyth:2003im, Lyth:2004gb}, where the homogeneous equations of motion are able to describe the inhomogeneous fields at leading order in a gradient expansion. This can be used to reconstruct the primordial density perturbation on super-Hubble scales, by combining stochastic inflation with the $\delta N$ formalism~\cite{Starobinsky:1982ee, Starobinsky:1986fxa, Sasaki:1995aw, Wands:2000dp, Lyth:2004gb}. The latter provides a framework in which curvature perturbations are related to fluctuations in the integrated amount of expansion and can be characterised from the knowledge of the stochastic dynamics of the fields driving inflation. 

This gives rise to the stochastic-$\delta N$ approach~\cite{Enqvist:2008kt, Fujita:2013cna, Vennin:2015hra}, which has been recently used to derive the full probability distribution of the primordial density field~\cite{Pattison:2017mbe,Ezquiaga:2019ftu}, finding large deviations from a Gaussian statistics in the nonlinear tail of the distribution. At the technical level, the main task of the stochastic-$\delta N$ program is to work out the distribution of first-passage times through the end-of-inflation surface, starting from some initial conditions. In fact, the problem that has been solved so far is the one when those initial conditions correspond to a fixed point $\boldmathsymbol{\Phi}$ in field space (here $\boldmathsymbol{\Phi}$ is a vector containing the values of all coarse-grained fields, and possibly of their conjugate momenta in the case where deviations from the slow-roll attractor are considered~\cite{Grain:2017dqa, Pattison:2019hef}). However, the amplitude of the fluctuations at a given length scale $\lambda$ are related to the statistics of the first-passage times from an initial condition that rather corresponds to when that scale crosses out the Hubble radius during inflation.

In regimes where quantum diffusion plays a negligible role, the fields driving inflation follow a classical, deterministic trajectory, so as a first approximation, there is a one-to-one correspondence between a scale $\lambda$, or equivalently, the Fourier wavenumber $k=2\pi/\lambda$ that is associated to it, and the value of the fields $\boldmathsymbol{\Phi}$ at the time when $k=aH$, \ie when $k$ crosses out the Hubble radius (where $a$ is the Friedmann-Lema\^itre-Robertson-Walker metric scale factor, $H=\dot{a}/a$ is the Hubble parameter, and a dot denotes derivation with respect to cosmic time). In other words, there exists a function  
\bea
\label{eq:classical:link:k:Phi}
\boldmathsymbol{\Phi}_\ucl(k)
\eea
that relates $k$ to a specific point in field space, which can be used as the initial condition for the first-passage time problem.

However, there are regimes where this approximation may not be sufficient. In particular, in regions of the inflationary potential giving rise to large cosmological fluctuations, which may be relevant for primordial black holes, the effect of quantum diffusion is important~\cite{Kawasaki:2015ppx, Clesse:2015wea, Pattison:2017mbe, Ezquiaga:2019ftu, Vennin:2020kng} and the classical trajectory cannot be used as a proxy for the typical realisations of the stochastic process. Furthermore, even if quantum diffusion plays an important role only at late time during inflation, and directly affects small scales only, the breakdown of \Eq{eq:classical:link:k:Phi} distorts the link between wavenumbers and field values at \emph{all} scales, which implies that a substantial stochastic imprint may be left on the large scales observed in the Cosmic Microwave Background (CMB), even if they emerge at a time when quantum diffusion is negligible. 

The aim of this paper is therefore to solve the first-passage time problem associated to a given scale rather than to a given field value, in order to bridge the final gap between the stochastic-$\delta N$ formalism and observable predictions. It is organised as follows. In \Sec{sec:stoch_deltaN_one}, we first review the stochastic-$\delta N$ formalism and explain how the first-passage time problem can be solved. In \Sec{sec:PS}, we present our calculation of the power spectrum in a generic inflating setup. In the quasi de-Sitter limit, we show that it boils down to deriving the probability distribution of the fields value at a given number of \efolds~before the end of inflation, and we explain how this can be computed from the solution of the Fokker-Planck and first passage-time problems. In \Sec{sec:SmallScales:Impact:LargeScales}, we apply our results to the case where quantum diffusion is efficient at small scales only, and we show that even then, it affects the power spectrum at large scales. We summarise our main results in \Sec{sec:conclusion} where we also present our conclusions, and we end the paper by \App{sec:FP:sol:Flat}, where some technical aspects of the calculation performed in \Sec{sec:SmallScales:Impact:LargeScales} are deferred. 

%%%%%%%%%%%%%%%%%%%%%%%%%%%%%%%%%%%%%%%%%%%%%%%%%%%%%
%%%%%%%%%%%%%%%%%%%%%%%%%%%%%%%%%%%%%%%%%%%%%%%%%%%%%
\section{The stochastic-$\delta N$ formalism and the first-passage time problem}
\label{sec:stoch_deltaN_one}
%%%%%%%%%%%%%%%%%%%%%%%%%%%%%%%%%%%%%%%%%%%%%%%%%%%%%
%%%%%%%%%%%%%%%%%%%%%%%%%%%%%%%%%%%%%%%%%%%%%%%%%%%%%
%
\subsection{The stochastic-$\delta N$ formalism}
\label{sec:Stoch:DeltaN:review}
On super Hubble scales, the curvature perturbation is related to the amount of expansion $N=\ln a$  from an initial flat space-time slice to a final space-time slice of uniform energy density $\rho$,
\bea
\label{eq:zeta:deltaN}
\zeta(t, \bm{x}) = N(t, \bm{x}) - \overline{N}(t) \equiv \delta N \, ,
\eea
where $\overline{N}(t)$ is the unperturbed number of \efolds. In the separate universe approach (the validity of which has recently been shown to extend beyond slow roll in \Refa{Pattison:2019hef}), on super-Hubble scales, gradients can be neglected, so $N(t, \bm{x})$ is the amount of expansion in unperturbed, homogeneous universes. The curvature perturbation $\zeta$ can thus be extracted from the knowledge of the durations of inflation in such universes. These durations vary under quantum fluctuations in the fields that drive inflation, which can be described in the stochastic inflation formalism by coarse-graining the fields above a fixed physical scale $k_\sigma=\sigma a H$, 
\bea
\hat{\boldmathsymbol{\Phi}}_{\mathrm{cg}}(\bm{x}) = \frac{1}{\left(2\pi\right)^{3/2}}\int_{k<k_\sigma} \dd^3 \bm{k} \hat{\boldmathsymbol{\Phi}}_{\bm{k}} \ee^{-i \bm{k}\cdot \bm{x}} ,
\eea
where $\sigma\ll 1$ is a fixed parameter that sets the scale at which quantum fluctuations backreact onto the local homogeneous geometry. At leading order in the gradient expansion, the quantum fields $\hat{\boldmathsymbol{\Phi}}_{\mathrm{cg}}$ can be described by classical random variables ${\boldmathsymbol{\Phi}}_{\mathrm{cg}}$, which obey stochastic Langevin equations of the form
\bea
\label{eq:Langevin}
\frac{\dd {\boldmathsymbol{\Phi}}_{\mathrm{cg}}}{\dd N} = \boldmathsymbol{F}_{\mathrm{cl}}\left(\boldmathsymbol{\Phi}_{\mathrm{cg}}\right)+\boldmathsymbol{\xi},
\eea
where $\boldmathsymbol{F}_{\mathrm{cl}}\left(\boldmathsymbol{\Phi}\right)$ encodes the classical equations of motion, and $\boldmathsymbol{\xi}$ is a white Gaussian noise with vanishing mean, and variance given by
\bea
\left\langle \xi_i\left(\bm{x},N_i\right) \xi_j\left(\bm{x},N_j\right) \right\rangle = \frac{\dd\ln k_\sigma}{\dd N} \calP_{\Phi_i,\Phi_j}\left[k_\sigma(N_i),N_i\right]\delta(N_i-N_j) ,
\eea
which describes the continuous inflow of modes into the coarse-grained sector.
In this expression, $\calP_{\Phi_i,\Phi_j}[k_\sigma(N),N]=k^3/(2\pi^2)\Rea\lbrace \Phi_i[k_\sigma(N)] \Phi_j^*[k_\sigma(N)]\rbrace$ is the reduced cross power spectrum between the field variables $\Phi_i$ and $\Phi_j$, evaluated at the scale $k_\sigma(N)$ and at time $N$. Note that when deriving the Langevin equation~(\ref{eq:Langevin}), the time coordinate (here the number of \efolds~$N$) has not been perturbed, so these power spectra must be evaluated in the gauge where $N$ is uniform, see \Refa{Pattison:2019hef}.

The Langevin equation~(\ref{eq:Langevin}) gives rise to a Fokker-Planck equation that drives the probability to find the field $\boldmathsymbol{\Phi}_{\mathrm{cg}}$ at position $\boldmathsymbol{\Phi}$ in field space at time $N$, given that it was at position $\boldmathsymbol{\Phi}_\uin$ at a previous time $N_\uin$,
\bea
\label{eq:Fokker:Planck}
\frac{\dd }{\dd N}P\left(\boldmathsymbol{\Phi},N\vert \boldmathsymbol{\Phi}_\uin,N_\uin\right) = \mathcal{L}_{\mathrm{FP}}\left(\boldmathsymbol{\Phi}\right)\cdot P\left(\boldmathsymbol{\Phi},N\vert \boldmathsymbol{\Phi}_\uin,N_\uin\right) .
\eea
In this expression, $ \mathcal{L}_{\mathrm{FP}}\left(\boldmathsymbol{\Phi}\right)$ is a differential operator of second order in phase space (\ie it contains first and second derivatives with respect to the field coordinates $\Phi_i$), called the Fokker-Planck operator.\footnote{When deriving \Eq{eq:Fokker:Planck} from \Eq{eq:Langevin}, one needs to specify the discretisation scheme with which \Eq{eq:Langevin} is interpreted. While this has important consequences for multiple-field systems in curved field spaces, where only the Stratonovitch scheme is compatible with field-space covariance~\cite{Pinol:2020cdp}, this has less importance for single-field systems. The generic formalism we construct in \Sec{sec:PS} does not specify the details of the Fokker-Planck operator and is therefore independent of the choice of discretisation scheme. For the specific model we later consider in \Sec{sec:SmallScales:Impact:LargeScales}, all discretisation schemes give the same result.\label{footnote:Discretisation:Strato:vs:Ito}}

The duration of inflation under the process~\eqref{eq:Langevin}, starting from the location $\boldmathsymbol{\Phi}_\uin$ in field space, is a random variable that we denote $\mathcal{N}$. Its probability distribution, $\Pfpt(\mathcal{N},\boldmathsymbol{\Phi}_\uin)$, can be shown to obey the adjoint Fokker-Planck equation~\cite{Vennin:2015hra, Pattison:2017mbe}
\bea
\label{eq:adjoint:FP}
\frac{\dd \Pfpt(\mathcal{N},\boldmathsymbol{\Phi}) }{\dd \mathcal{N} }= \mathcal{L}_{\mathrm{FP}}^\dagger \left(\boldmathsymbol{\Phi}\right)\cdot \Pfpt(\mathcal{N},\boldmathsymbol{\Phi}) ,
\eea
where $\mathcal{L}_{\mathrm{FP}}^\dagger(\boldmathsymbol{\Phi})$ is the adjoint Fokker-Planck operator, related to the Fokker-Planck operator via $\int \dd \boldmathsymbol{\Phi} F_1(\boldmathsymbol{\Phi}) \mathcal{L}_{\mathrm{FP}}(\boldmathsymbol{\Phi})\cdot F_2(\boldmathsymbol{\Phi}) = \int \dd \boldmathsymbol{\Phi} F_2(\boldmathsymbol{\Phi}) \mathcal{L}_{\mathrm{FP}}^\dagger(\boldmathsymbol{\Phi})\cdot F_1(\boldmathsymbol{\Phi}) $. Since \Eq{eq:adjoint:FP} is a partial differential equation, it needs to be solved with some boundary condition. The first condition is given by the end-of-inflation surface $\mathcal{C}_\uend=\lbrace \boldmathsymbol{\Phi},\rho(\boldmathsymbol{\Phi})=\rho_\uend \rbrace$,\footnote{In single-field inflation, assuming that the slow-roll attractor has been reached before the end of inflation, the field and its conjugated momentum are directly related, so inflation ends through a single point in field phase space, and $\rho_\uend$ can be taken as the energy density at that point. The situation is more involved in multiple-field systems, but assuming that quantum diffusion becomes negligible when inflation terminates, the choice of the precise value for $\rho_\uend$ becomes irrelevant.} where the number of inflationary \efolds~necessarily vanishes, hence $\Pfpt(\mathcal{N},\boldmathsymbol{\Phi}) =\delta(\mathcal{N})$ for $\boldmathsymbol{\Phi}\in\mathcal{C}_\uend$. In hilltop models, $\mathcal{C}_\uend$ makes the inflating domain of field space compact and is enough to fully specify the problem. Otherwise, another boundary condition has to be added, which in practice we take to be a reflective boundary condition high enough in the potential~\cite{Assadullahi:2016gkk, Vennin:2016wnk}, along some surface $\mathcal{C}_{\mathrm{uv}}$, such that $[\bm{u}(\boldmathsymbol{\Phi})\cdot\bm{\nabla}]P_{\boldmathsymbol{\Phi}}\left(\mathcal{N}\right)=0$ when $\boldmathsymbol{\Phi} \in \mathcal{C}_{\mathrm{uv}}$, where $\bm{u}$ is a field-space vector orthogonal to $\mathcal{C}_{\mathrm{uv}}$. The details of this second boundary condition are anyway irrelevant as long as inflation proceeds at sub-Planckian energies~\cite{Assadullahi:2016gkk, Vennin:2016wnk}. 

From \Eq{eq:adjoint:FP}, the statistics of $\mathcal{N}$, hence of $\delta\mathcal{N}=\mathcal{N}-\left\langle \mathcal{N} \right\rangle$  (where $\langle \cdot \rangle$ denotes statistical average), and in turn the one of $\zeta$ through \Eq{eq:zeta:deltaN}, can be inferred. This is the stochastic-$\delta N$ program.
Since the power spectrum corresponds to the two-point function of the curvature perturbation, in practice, one needs to compute the two first moments of $\mathcal{N}$, given that $\langle \delta \mathcal{N}^2\rangle = \langle \mathcal{N}^2\rangle - \langle \mathcal{N} \rangle^2$. By applying the adjoint Fokker-Planck operator to the definition of the $n^\mathrm{th}$ moment of $\mathcal{N}$, $\langle \mathcal{N}^n\rangle(\boldmathsymbol{\Phi}) = \int \Pfpt(\mathcal{N},\boldmathsymbol{\Phi}) \mathcal{N}^n \dd \mathcal{N}$, after integration by part, one obtains the iterative set of differential equations~\cite{Vennin:2015hra}
\bea
\label{eq:adjoint:FP:moment}
\mathcal{L}_{\mathrm{FP}}^\dagger \left(\boldmathsymbol{\Phi}\right)\cdot \langle \mathcal{N}^n\rangle(\boldmathsymbol{\Phi}) = -n \langle \mathcal{N}^{n-1}\rangle(\boldmathsymbol{\Phi}) .
\eea
Starting from $\langle \mathcal{N}^0\rangle =1$, one can then solve iteratively for $\langle \mathcal{N}\rangle $ and $\langle \mathcal{N}^2\rangle$.

Let us stress that $\Pfpt(\mathcal{N},\boldmathsymbol{\Phi})$ corresponds to the distribution of $\mathcal{N}$ associated to a given initial point $\boldmathsymbol{\Phi}$ in field space. However, as explained in \Sec{sec:intro}, in order to make contact with observations, one should determine the distribution of $\mathcal{N}$ associated to a given scale $k$. The link between $\boldmathsymbol{\Phi}$ and $k$ thus remains to be specified, and this is the goal of the present paper.
\subsection{Power spectrum in the low-diffusion limit}
In the regime where the stochastic noise plays a negligible role in the Langevin equation~\eqref{eq:Langevin}, at first order in the noise, all realisations follow the classical, deterministic equation of motion. The value of the fields $\boldmathsymbol{\Phi}$ at the time when the scale $k$ crosses out the Hubble radius, \ie when $k=aH$, is therefore the same for all realisations, which defines the function $\boldmathsymbol{\Phi}_\ucl(k)$ introduced in \Eq{eq:classical:link:k:Phi}, which can be inverted into the function $k_\ucl(\boldmathsymbol{\Phi})$. 

At this order, the fluctuations in the duration of inflation starting from $\boldmathsymbol{\Phi}_\uin$ receive a contribution from all modes that cross out the Hubble radius between the time when $\boldmathsymbol{\Phi}=\boldmathsymbol{\Phi}_\uin$ and the end of inflation, \ie from all modes $k$ such that $k_\uc(\boldmathsymbol{\Phi}_\uin)<k<k_\uend$, where $k_\uend$ corresponds to the Hubble scale at the end of inflation. One thus has~\cite{Fujita:2013cna}
\bea
\left\langle \delta\mathcal{N}^2(\boldmathsymbol{\Phi})\right\rangle = \int_{k_\ucl(\boldmathsymbol{\Phi})<k<k_\uend}\calP_\zeta(k) \dd\ln k .
\eea
A detailed derivation of this (otherwise intuitive) formula will be provided in \Sec{sec:PS}, but for now, let us note that by differentiating both hands with respect to the field-space coordinate along the background, classical trajectory, one obtains
\bea
\label{eq:PowerSpectrum:Low:Diffusion:Limit}
\calP_\zeta(k)=-\frac{\dd\left\langle \delta \mathcal{N}^2\right\rangle\left[\boldmathsymbol{\Phi}_\ucl(k)\right]}{\dd\ln(k)}\, .
\eea
\subsection{Case of a slowly-rolling single field}
For illustrative purpose, let us apply the above calculational program to the case of a single field in the slow-roll regime, and see how the standard formula for the power spectrum can be recovered. For a single scalar field $\phi$ with a canonical kinetic term and potential function $V(\phi)$, in the slow-roll regime, the Fokker-Planck and adjoint Fokker-Planck operators are respectively given by~\cite{Starobinsky:1986fx, Starobinsky:1994bd}
\bea
\label{FP_operator_1dim}
\frac{1}{\Mp^2}\mathcal{L}_{\mathrm{FP}}(\phi)\cdot F &=&\frac{\partial}{\partial\phi}\left[\frac{v'(\phi)}{v(\phi)} F(\phi) \right]+\frac{\partial^2}{\partial\phi^2}\left[v(\phi) F(\phi) \right]\\
\frac{1}{\Mp^2}\mathcal{L}_{\mathrm{FP}}^\dagger(\phi)\cdot F  &=&-\frac{v'(\phi)}{v(\phi)}\frac{\partial}{\partial\phi} F(\phi)  +v\frac{\partial^2}{\partial\phi^2} F(\phi) \, ,
\eea
where $F$ is a dummy function on which the differential operators act, $\Mp$ is the reduced Planck mass, a prime denotes derivation with respect to the field value $\phi$, and we have introduced the reduced potential
\bea
v(\phi)=\frac{V(\phi)}{24\pi^2\Mp^4}
\eea
for convenience. This allows one to integrate \Eq{eq:adjoint:FP:moment} exactly, and one obtains~\cite{Vennin:2015hra, Assadullahi:2016gkk}
\bea
\label{eq:moments:SingleField:SlowRoll}
\left\langle \mathcal{N}^n \right\rangle (\phi) =n  \int_{\phi_\uend}^\phi \frac{\dd\phi_1}{\Mp} \int_{\phi_1}^{\phi_\mathrm{uv}} \frac{\dd\phi_2}{\Mp} \frac{\left\langle \mathcal{N}^{n-1} \right\rangle \left(\phi_2\right) }{v\left(\phi_2\right)} \exp\left[\frac{1}{v\left(\phi_2\right)}-\frac{1}{v\left(\phi_1\right)}\right] ,
\eea
where $\phi_\uend$ is the value of $\phi$ at the end of inflation, such that $\langle \mathcal{N}^n \rangle (\phi_\uend)=0$, and the upper bound of the second integral is set to $\phi_\mathrm{uv}$ in order to satisfy the second boundary condition, namely $\langle \mathcal{N}^n \rangle' (\phi_\mathrm{uv})=0$. This formula allows one to compute all moments iteratively, for any potential function $v(\phi)$.

Let us now consider the low-diffusion limit.
Since $v$ measures the potential energy in Planckian units, $1/v\gg 1$ in the regime of interest, which allows one to perform a saddle-point expansion of the integrals appearing in \Eq{eq:moments:SingleField:SlowRoll}, owing to the large exponential terms. Under the condition $\left\vert v''  \right\vert v^2\ll {v'}^2$ (which imposes that the slope of the potential is large enough so that the classical drift dominates over quantum diffusion), one obtains~\cite{Vennin:2015hra}
\bea
\label{eq:meanN:class}
\left\langle \mathcal{N}\right\rangle (\phi) &\simeq& \int_{\phi_\uend}^\phi \frac{\dd\bar{\phi}}{\Mp^2}\frac{v\left(\bar{\phi}\right)}{v'\left(\bar{\phi}\right)}\\
\left\langle \delta \mathcal{N}^2\right\rangle (\phi) &\simeq& 2 \int_{\phi_\uend}^\phi \frac{\dd\bar{\phi}}{\Mp^4}\frac{v^4\left(\bar{\phi}\right)}{{v'}^3\left(\bar{\phi}\right)}\, .
\label{eq:meandeltaN2:class}
\eea
At leading order in slow roll where $H$ is almost constant, $k/k_\uend = a_\ucl(k)H_\ucl(k)/(a_\uend H_\uend)\simeq e^{N_\ucl(k)-N_\uend}$, hence the derivation with respect to $\ln(k)$ in \Eq{eq:PowerSpectrum:Low:Diffusion:Limit} is a derivation with respect to $N_\ucl(k)-N_\uend \simeq - \langle \mathcal{N}\rangle [\phi_\ucl(k)]$, which gives rise to
\bea
\label{eq:Pzeta:classical:SingleField:SlowRoll}
\calP_\zeta(k)\simeq\frac{\left\langle \delta\mathcal{N}^2\right\rangle'\left[\phi_\ucl(k)\right]}{\left\langle\mathcal{N}\right\rangle'\left[\phi_\ucl(k)\right]}\simeq \frac{2}{\Mp^2}\frac{v^3\left[\phi_\ucl(k)\right]}{{v'}^2\left[\phi_\ucl(k)\right]}\, .
\eea
In the last equality, we have made use of \Eqs{eq:meanN:class} and~\eqref{eq:meandeltaN2:class}. In this way one recovers the standard formula for the power spectrum of curvature perturbations in single-field slow-roll inflation~\cite{Mukhanov:1985rz, Mukhanov:1988jd}. An important remark is that the validity of this formula relies on assuming quantum diffusion to be low not only at the time when the observed scales are produced, but at any later time during inflation. While the former condition is known to be satisfied for the scales observed in the CMB, the later condition is not guaranteed, and is even explicitly violated in models giving rise to large cosmological perturbations at small scales, in particular those leading to primordial black holes. The goal of the rest of this paper is therefore to go beyond those approximations.
%%%%%%%%%%%%%%%%%%%%%%%%%%%%%%%%%%%%%%%%%%%%%%%%%%%%%
%%%%%%%%%%%%%%%%%%%%%%%%%%%%%%%%%%%%%%%%%%%%%%%%%%%%%
\section{Power spectrum in stochastic inflation}
\label{sec:PS}
%%%%%%%%%%%%%%%%%%%%%%%%%%%%%%%%%%%%%%%%%%%%%%%%%%%%%
%%%%%%%%%%%%%%%%%%%%%%%%%%%%%%%%%%%%%%%%%%%%%%%%%%%%%

\subsection{Encoding spatial correlations into statistical trees}
\label{sec:StatisticalTree}
\begin{figure}[t]
\begin{center}
\includegraphics[width=1.\textwidth]{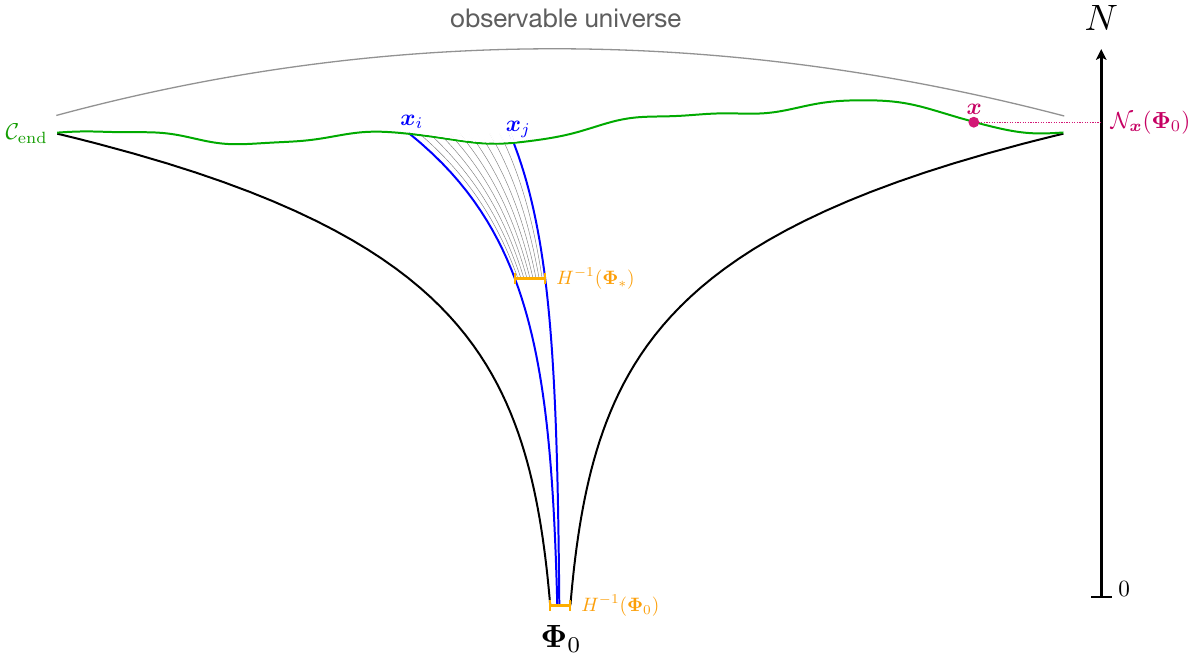}
\caption{Space-time diagram (here in one dimension for display convenience) during inflation. For two comoving points labeled by $\bm{x}_i$ and $\bm{x}_j$ on the end-of-inflation surface, there exists a time at which their distance equals the Hubble radius. Prior to this time, they belong to the same Hubble patch and their dynamics is identical. Past this time, their evolutions become statistically independent, and the time that separates this point from the end-of-inflation surface is related to the spatial distance between $\bm{x}_i$ and $\bm{x}_j$.}  
\label{fig:sketch3}
\end{center}
\end{figure}
Since the power spectrum is nothing but the Fourier transform of the two-point correlation function in real space, one must be able to describe the spatial structure of the correlations between the durations of inflation at different points in order to compute the power spectrum. However, since the Langevin equation describes the dynamics of independent patches of the universe, and a priori carries no information about their relative spatial positions, one may be concerned that this spatial structure is lost and that only one-point correlation functions can be computed in stochastic inflation. In fact, as we shall now see, the distance between two patches is encoded in the time at which they become statistically independent, such that the two-point function is contained in the one-point dynamics of the Langevin equation. 

More explicitly, the situation we consider is depicted in the space-time diagram of \Fig{fig:sketch3}. During inflation, there is a point before which the entire observable universe lies within a single Hubble patch. We call $\boldmathsymbol{\Phi}_0$ the value of the inflationary fields inside that patch (since $\boldmathsymbol{\Phi}$ is coarse-grained on super-Hubble scales, it is indeed homogeneous within a Hubble patch). Each comoving point on that patch, labeled by its position $\boldmathsymbol{x}$, crosses the end surface (displayed in green) at a different time, that we denote $\mathcal{N}_{\boldmathsymbol{x}}(\boldmathsymbol{\Phi}_0)$ (see the violet point in \Fig{fig:sketch3}).

Let us consider two points $\bm x_{i}$ and $\bm x_{j}$ on the end-of-inflation surface. They are the end points of two comoving lines displayed in blue. Prior to the end of inflation, there is a time when the distance between these two lines coincides with the Hubble radius, and we call $\boldmathsymbol{\Phi}_*$ the value of the fields at that point (since $\boldmathsymbol{\Phi}$ is uniform across a Hubble patch, it is indeed the same at $\bm x_{i}$ and $\bm x_{j}$ at that time). This defines the function
\bea
\boldmathsymbol{\Phi}_*\left(\bm{x}_i,\bm{x}_j\right).
\eea
Before that point, the two blue lines lie within the same Hubble patch, hence they share the same history, \ie they are described by the same realisation of the Langevin equation. Past that ``splitting'' point, owing to the Markovian property of the stochastic process we consider, they become statistically independent. If $\overline{N}(\boldmathsymbol{\Phi}_0)$ denotes the mean value of $\mathcal{N}_{\bm{x}}(\boldmathsymbol{\Phi}_0)$ over all end points, according to \Eq{eq:zeta:deltaN}, one has
\bea
\label{eq:zeta:Ni}
\zeta(\bm{x}_i)=\mathcal{N}_i(\boldmathsymbol{\Phi}_0)-\overline{N}(\boldmathsymbol{\Phi}_0)\, ,
\eea
where $\mathcal{N}_i$ is a short-hand notation for $\mathcal{N}_{\bm{x}_i}$. One can also introduce the physical distance $r$ between the two end points. On the splitting patch (displayed in orange and labeled with $H^{-1}(\boldmathsymbol{\Phi}_*)$ in \Fig{fig:sketch3}), let $\boldmathsymbol{y}$ label the comoving position along the geodesic connecting $\boldmathsymbol{x}_i$ and $\boldmathsymbol{x}_j$. Each such point gives rise to one of the grey lines in \Fig{fig:sketch3}, and undergoes a number of inflationary \efolds~that we denote $\mathcal{N}_{\boldmathsymbol{y}}[\boldmathsymbol{\Phi}_*(\bm{x}_i,\bm{x}_j)]$. The physical distance between $\boldmathsymbol{x}_i$ and $\boldmathsymbol{x}_j$ on the end surface is thus given by
\bea
\label{eq:distance:def}
r\left(\bm{x}_i,\bm{x}_j\right)=\int_0^{H^{-1}\left[\boldmathsymbol{\Phi}_*(\bm{x}_i,\bm{x}_j)\right]} \ee^{\mathcal{N}_{\boldmathsymbol{y}}[\boldmathsymbol{\Phi}_*(\bm{x}_i,\bm{x}_j)]} \dd
y\, ,
%\boldmathsymbol{y}\, .
\eea
where we have introduced $\boldmathsymbol{y}=\boldmathsymbol{x}_i+y (\boldmathsymbol{x}_j-\boldmathsymbol{x}_i)/\vert \boldmathsymbol{x}_j-\boldmathsymbol{x}_i\vert$.
The two-point function of $\zeta$ at separation $r$ is then given by the expectation value of the product $\zeta(\bm{x}_i) \zeta(\bm{x}_j)$, averaged over all pairs of comoving lines that share the same distance $r$. The power spectrum is then simply defined as the Fourier transform of that function, such that 
\bea
\label{eq:Power:Spectrum:Exact:CardinalSine}
\left\langle \zeta(\bm{x}_i) \zeta(\bm{x}_j) \right\rangle_{r\left(\bm{x}_i,\bm{x}_j\right)=\tilde{r}}
 = \int_0^\infty 
\calP_\zeta(k)
\frac{\sin(k \tilde{r})}{k \tilde{r}} \frac{\dd k}{k}.
\eea
A remark is in order regarding the appearance of the cardinal sine function in this expression. In principle, the two-point function of the noise $\xi$ in the Langevin equation~\eqref{eq:Langevin} also depends through a cardinal sine function on the distance (in Hubble units) between the two points at which it is evaluated~\cite{Starobinsky:1994bd}. Therefore, the realisations of the noise at two distant points are still slightly correlated after they split (\ie after they no longer belong to the same Hubble patch), and are not exactly independent. By neglecting this remaining amount of correlation, one is effectively approximating the cardinal sine function with a Heaviside function. This approximation is well justified since those correlations quickly decay on super-Hubble scales (and given that their details anyway depend on the window function that has been used to coarse grain the fields), but for consistency, the cardinal sine function in \Eq{eq:Power:Spectrum:Exact:CardinalSine} should also be replaced by a Heaviside function. By differentiating both hands of \Eq{eq:Power:Spectrum:Exact:CardinalSine} with respect to $\tilde{r}$, one then obtains
\bea
\label{eq:Power:Spectrum:Exact}
\calP_\zeta(k) 
=
- \left.\frac{\partial }{\partial\ln(\tilde{r})} \left\langle \zeta(\bm{x}_i) \zeta(\bm{x}_j) \right\rangle_{r\left(\bm{x}_i,\bm{x}_j\right)=\tilde{r}}
\right\vert_{\tilde{r}=1/k}.
\eea
This formula can be used to extract the power spectrum from a numerical lattice simulation of the Langevin equations.
\subsection{Using the first-passage-time moments}
Let us now try to benefit from the results recalled in \Sec{sec:Stoch:DeltaN:review}, which provide the statistical moments of the first passage time through the end surface from a given point in field space. In order to make use of these formulas, it is convenient to introduce $P_r(\boldmathsymbol{\Phi_*})$, which is the probability density associated to the fields value in the splitting patch, for two comoving lines separated by $r$ on the end surface. The calculation of this function will be discussed below. 

Along the line labeled by $\bm{x}_i$, let us first write $\mathcal{N}_i(\boldmathsymbol{\Phi}_0)$ as the sum of the number of \efolds~realised between the original patch and the splitting patch, which we denote $\mathcal{N}_i[\boldmathsymbol{\Phi}_0 \to \boldmathsymbol{\Phi}_*(\bm{x}_i,\bm{x}_j)]$ (and which is not necessarily a \emph{first}-passage time), and the number of \efolds~realised between the splitting patch and the end surface, $\mathcal{N}_i[\boldmathsymbol{\Phi}_*(\bm{x}_i,\bm{x}_j)]$, such that \Eq{eq:zeta:Ni} gives rise to
\bea
\label{eq:zeta:Ni:expanded}
\zeta\left(\bm{x}_i\right)=\mathcal{N}_i\left[\boldmathsymbol{\Phi}_0 \to \boldmathsymbol{\Phi}_*\left(\bm{x}_i,\bm{x}_j\right)\right]+\mathcal{N}_i\left[\boldmathsymbol{\Phi}_*\left(\bm{x}_i,\bm{x}_j\right)\right]-\overline{N}\left(\boldmathsymbol{\Phi}_0\right)\, .
\eea
Obviously, a similar formula holds for $\zeta(\bm{x}_j)$, and given that the two branches $\bm{x}_i$ and $\bm{x}_j$ share the same common past before the splitting patch, one has $\mathcal{N}_i[\boldmathsymbol{\Phi}_0 \to \boldmathsymbol{\Phi}_*(\bm{x}_i,\bm{x}_j)]=\mathcal{N}_j[\boldmathsymbol{\Phi}_0 \to \boldmathsymbol{\Phi}_*(\bm{x}_i,\bm{x}_j)]$. Because of \Eq{eq:Power:Spectrum:Exact}, one must cross correlate \Eq{eq:zeta:Ni:expanded} with the same expression when $\bm{x}_i$ is replaced by $\bm{x}_j$, under the condition that the distance between $\bm{x}_i$ and $\bm{x}_j$ is fixed. This gives rise to several terms that we now discuss one by one.

Let us first consider the product term $\mathcal{N}_i(\boldmathsymbol{\Phi}_*)\mathcal{N}_j(\boldmathsymbol{\Phi}_*)$ (hereafter, for notational convenience, we drop the arguments of the function $\boldmathsymbol{\Phi}_*$). Using the chain rule for conditional probabilities, one has
\bea
\label{eq:chain:rule}
\kern-1em
\left\langle  \mathcal{N}_i\left(\boldmathsymbol{\Phi}_*\right)\mathcal{N}_j\left(\boldmathsymbol{\Phi}_*\right) \right\rangle_{r\left(\bm{x}_i,\bm{x}_j\right)=\tilde{r}} = 
\int \dd\widetilde{\boldmathsymbol{\Phi}}_* P_{\tilde{r}}\left(\widetilde{\boldmathsymbol{\Phi}}_*\right)
\left\langle  \mathcal{N}_i\left(\boldmathsymbol{\Phi}_*\right)\mathcal{N}_j\left(\boldmathsymbol{\Phi}_*\right) \right\rangle_{r\left(\bm{x}_i,\bm{x}_j\right)=\tilde{r}, \boldmathsymbol{\Phi}_*\left(\bm{x}_i,\bm{x}_j\right)=\widetilde{\boldmathsymbol{\Phi}}_*}\, .
\eea
An important remark is that if the distance between $\boldmathsymbol{x}_i$ and $\boldmathsymbol{x}_j$ is much larger than the Hubble radius on the end surface (which needs to be the case since the stochastic formalism only allows us to describe long-distance correlations), most lines labeled by $\boldmathsymbol{y}$ (and displayed in grey in \Fig{fig:sketch3}) split from the ones ending in $\boldmathsymbol{x}_i$ and $\boldmathsymbol{x}_j$ much before the end of inflation. As a consequence, the number of \efolds~realised along these lines is almost independent from the one undergone by the lines ending in $\boldmathsymbol{x}_i$ and $\boldmathsymbol{x}_j$, hence \Eq{eq:distance:def} indicates that the value of $r$ is (almost) uncorrelated with $\mathcal{N}_i\left(\boldmathsymbol{\Phi}_*\right)$ or $\mathcal{N}_j\left(\boldmathsymbol{\Phi}_*\right)$. This implies that $\langle  \mathcal{N}_i(\boldmathsymbol{\Phi}_*)\mathcal{N}_j(\boldmathsymbol{\Phi}_*) \rangle_{r(\bm{x}_i,\bm{x}_j)=\tilde{r}, \boldmathsymbol{\Phi}_*(\bm{x}_i,\bm{x}_j)=\widetilde{\boldmathsymbol{\Phi}}_*} = \langle  \mathcal{N}_i(\boldmathsymbol{\Phi}_*)\mathcal{N}_j(\boldmathsymbol{\Phi}_*) \rangle_{\boldmathsymbol{\Phi}_*(\bm{x}_i,\bm{x}_j)=\widetilde{\boldmathsymbol{\Phi}}_*}=\langle \mathcal{N}\rangle^2(\widetilde{\boldmathsymbol{\Phi}}_*)$, and one recovers the first statistical moment of the first-passage time computed in \Sec{sec:Stoch:DeltaN:review}.

The chain rule of \Eq{eq:chain:rule} applies more generally to any function of $\bm{x}_i$ and $\bm{x}_j$, so the other terms can also be evaluated as integrals over $\boldmathsymbol{\Phi}_*$ of correlators at fixed $r$ and $\boldmathsymbol{\Phi}_*$. For the term $\mathcal{N}_i(\boldmathsymbol{\Phi}_0\to \boldmathsymbol{\Phi}_*)\mathcal{N}_j(\boldmathsymbol{\Phi}_0\to \boldmathsymbol{\Phi}_*)$, since $r$ depends only on the post-splitting-patch dynamics, the fact that $r$ is fixed is irrelevant and one has 
\bea
\langle \mathcal{N}_i\left(\boldmathsymbol{\Phi}_0\to \boldmathsymbol{\Phi}_*\right) \mathcal{N}_j\left(\boldmathsymbol{\Phi}_0\to \boldmathsymbol{\Phi}_*\right) \rangle_{r\left(\bm{x}_i,\bm{x}_j\right)=\tilde{r}, \boldmathsymbol{\Phi}_*\left(\bm{x}_i,\bm{x}_j\right)=\widetilde{\boldmathsymbol{\Phi}}_*}=\left\langle \mathcal{N}^2\left(\boldmathsymbol{\Phi}_0\to \widetilde{\boldmathsymbol{\Phi}}_*\right)\right\rangle 
\eea
where we have used that $\mathcal{N}_i(\boldmathsymbol{\Phi}_0\to \boldmathsymbol{\Phi}_*)=\mathcal{N}_j(\boldmathsymbol{\Phi}_0\to \boldmathsymbol{\Phi}_*)$ as mentioned above. Let us stress that contrary to the previous term, this quantity is not a statistical moment of a first passage time, so it cannot be evaluated with the techniques presented in \Sec{sec:Stoch:DeltaN:review}, but we will see below that it cancels out with other contributions in the final result.

 The Markovian nature of the process also implies that $\mathcal{N}_j(\boldmathsymbol{\Phi}_*)$ and $\mathcal{N}_i(\boldmathsymbol{\Phi}_0 \to \boldmathsymbol{\Phi}_*)=\mathcal{N}_j(\boldmathsymbol{\Phi}_0\rightarrow\boldmathsymbol{\Phi}_*)$ are independent when $\boldmathsymbol{\Phi}_*$ is fixed, so  
\bea
\left\langle \mathcal{N}_i\left(\boldmathsymbol{\Phi}_0 \to \boldmathsymbol{\Phi}_*\right) \mathcal{N}_j\left(\boldmathsymbol{\Phi}_*\right) \right\rangle_{r\left(\bm{x}_i,\bm{x}_j\right)=\tilde{r}, \boldmathsymbol{\Phi}_*\left(\bm{x}_i,\bm{x}_j\right)=\widetilde{\boldmathsymbol{\Phi}}_*}=  \left\langle \mathcal{N}(\boldmathsymbol{\Phi}_0 \to \widetilde{\boldmathsymbol{\Phi}}_*) \right\rangle\left\langle \mathcal{N}\right\rangle(\widetilde{\boldmathsymbol{\Phi}}_*)\, ,
\eea
where the last term is the first moment of a first-passage time. 

Those considerations allow one to compute the 9 terms that appear in the expectation value of $\zeta(\bm{x}_i) \zeta(\bm{x}_j)$, and after some rearrangements one obtains
\bea
\left\langle \zeta(\bm{x}_i) \zeta(\bm{x}_j) \right\rangle_{r\left(\bm{x}_i,\bm{x}_j\right)=\tilde{r}} &=& \int \dd{\boldmathsymbol{\Phi}}_* P_{\tilde{r}}\left(\boldmathsymbol{\Phi}_*\right)
 \left[
\left\langle \mathcal{N}^2\left(\boldmathsymbol{\Phi}_0 \to \boldmathsymbol{\Phi}_*\right)\right\rangle
+\left\langle \mathcal{N}\right\rangle^2\left({\boldmathsymbol{\Phi}}_*\right)
+\overline{N}^2\left(\boldmathsymbol{\Phi}_0\right)
 \right.  \nonumber \\  & & \kern-10em \left.
+2 \left\langle \mathcal{N}\left(\boldmathsymbol{\Phi}_0 \to \boldmathsymbol{\Phi}_*\right)\right\rangle
\left\langle \mathcal{N}\right\rangle\left({\boldmathsymbol{\Phi}}_*\right)
-2 \overline{N}\left(\boldmathsymbol{\Phi}_0\right)
\left\langle \mathcal{N}\left(\boldmathsymbol{\Phi}_0 \to \boldmathsymbol{\Phi}_*\right)\right\rangle
-2 \overline{N}\left(\boldmathsymbol{\Phi}_0\right)
\left\langle \mathcal{N}\right\rangle\left({\boldmathsymbol{\Phi}}_*\right)
\right]\, .
\eea
This expression can be further simplified by noting that, if one decomposes $\mathcal{N}(\boldmathsymbol{\Phi}_0) =  \mathcal{N}\left(\boldmathsymbol{\Phi}_0 \to \boldmathsymbol{\Phi}_*\right) + \mathcal{N}(\boldmathsymbol{\Phi}_*)$ and takes the stochastic average, one has $\overline{N}\left(\boldmathsymbol{\Phi}_0\right) = \langle \mathcal{N}(\boldmathsymbol{\Phi}_0 \to \boldmathsymbol{\Phi}_*)\rangle+\langle \mathcal{N}\rangle({\boldmathsymbol{\Phi}}_*)$, so by replacing $\overline{N}(\boldmathsymbol{\Phi}_0) $ in the above expression, it reduces to
\bea
\label{eq:PowerSpectrum:interm}
\left\langle \zeta(\bm{x}_i) \zeta(\bm{x}_j) \right\rangle_{r\left(\bm{x}_i,\bm{x}_j\right)=\tilde{r}} &=& 
\int \dd{\boldmathsymbol{\Phi}}_* P_{\tilde{r}}\left(\boldmathsymbol{\Phi}_*\right)
\left\langle \delta\mathcal{N}^2\left(\boldmathsymbol{\Phi}_0 \to \boldmathsymbol{\Phi}_*\right) \right\rangle\, .
\eea
Finally, let us note that, invoking again the Markovian nature of the process, $\mathcal{N}(\boldmathsymbol{\Phi}_0 \to \boldmathsymbol{\Phi}_*)$ and $\mathcal{N}(\boldmathsymbol{\Phi}_*)$ are two independent variables (when $\boldmathsymbol{\Phi}_*$ is fixed). Therefore by squaring the relation $\mathcal{N}(\boldmathsymbol{\Phi}_0) =  \mathcal{N}\left(\boldmathsymbol{\Phi}_0 \to \boldmathsymbol{\Phi}_*\right) + \mathcal{N}(\boldmathsymbol{\Phi}_*)$ given above, and after taking its stochastic average, one obtains $\langle \delta \mathcal{N}^2\left(\boldmathsymbol{\Phi}_0 \to \boldmathsymbol{\Phi}_*\right)\rangle= \langle \delta\mathcal{N}^2\rangle (\boldmathsymbol{\Phi}_0)-\langle \delta\mathcal{N}^2\rangle (\boldmathsymbol{\Phi}_*)$. Since the first term in the right-hand side of this expression does not depend on $\boldmathsymbol{\Phi}_*$, it provides a contribution proportional to $\int\dd\boldmathsymbol{\Phi}_* P_{\tilde{r}}(\boldmathsymbol{\Phi}_* )=1$ in \Eq{eq:PowerSpectrum:interm}, \ie a term that does not depend on $\tilde{r}$ so \Eq{eq:Power:Spectrum:Exact} gives rise to
\bea
\label{eq:Power:Spectrum:Exact:FPT}
\calP_\zeta(k) = \frac{1}{k} \int \dd{\boldmathsymbol{\Phi}}_* \left. \frac{\partial P_{r}\left(\boldmathsymbol{\Phi}_*\right)}{\partial r}\right\vert_{r=1/k} 
\left\langle \delta \mathcal{N}^2\right\rangle\left(\boldmathsymbol{\Phi}_*\right)\, .
\eea
This expression has the advantage to directly involve the second moment of the first-passage time, studied in \Sec{sec:Stoch:DeltaN:review}.  

\subsection{Field value at the splitting patch}
The next step in the calculation is to compute the probability associated to the field value at the splitting patch, $P_{r}\left(\boldmathsymbol{\Phi}_*\right)$, which appears in \Eq{eq:Power:Spectrum:Exact:FPT}. 

An important remark is that if two points are separated by a distance $r$ on the end surface, this distance should be red-shifted (or rather blue-shifted) to previous times using the backward \efold~number, $N_{\mathrm{bw}}=N_\uend-N$, \ie the number of \efolds~before the end of inflation. Indeed, as argued in \Refa{Tada:2016pmk}, observable quantities should be stated in terms of physical scales as seen by local observers, which imposes to label scales with backwards \efolds. Note that since the stochastic noise is turned off at the end of inflation, physical scales on the end surface are directly mapped to scales measured in observations today. 

As a consequence, if two comoving lines are separated by a distance $r$ on the end surface, they become independent when
\bea
\label{eq:Nbw:r:H(phistar)}
\ee^{N_{\mathrm{bw},*}}=r H\left(\boldmathsymbol{\Phi}_*\right)\, .
\eea
Along each realisation of the Langevin equation, one can determine when this condition is satisfied, record the value of $\boldmathsymbol{\Phi}_*$ at that time, and then reconstruct the probability distribution associated to $\boldmathsymbol{\Phi}_*$. As will be shown below, in the case of a one-dimensional field phase space (\ie for single-field systems that have reached the slow-roll attractor\footnote{Single-field setups that violate slow-roll for a transient period (and typically enter a phase of ultra slow roll) are usually preceded by a phase of slow roll inflation, in which case the dependence on initial condition is also erased.}), the result does not depend on the initial condition $\boldmathsymbol{\Phi}_0$, because of the Markovian nature of the process. In multiple field systems however, the distribution $P_{r}\left(\boldmathsymbol{\Phi}_*\right)$ does a priori depend on the choice of $\boldmathsymbol{\Phi}_0$ (or more generally, on the distribution function associated to $\boldmathsymbol{\Phi}_0$), which is a fundamental difference.

This prescription allows one to evaluate $P_{r}(\boldmathsymbol{\Phi}_*)$ numerically in a straightforward way, so the above considerations provide an explicit procedure to evaluate the power spectrum in stochastic inflation, which was one of the main goals of this paper.

In order to gain further analytical insight, let us notice that in the quasi de-Sitter limit where $H$ is almost a constant, $H(\boldmathsymbol{\Phi}_*)$ can be simply replaced by $H_\uend$ in \Eq{eq:Nbw:r:H(phistar)}, which allows one to define
\bea
\label{eq:Nbw(k):deSitter}
N_\mathrm{bw}(r)=\ln\left(r H_\uend \right)=-\ln\left(\frac{k}{k_\uend}\right)\, .
\eea
In this limit, the probability $P_{r}(\boldmathsymbol{\Phi}_*)$ becomes the one associated to the fields value at a fixed backward \efold~number, 
\bea
P_{r}\left(\boldmathsymbol{\Phi}_*\right)\simeq P_\mathrm{bw}\left[\boldmathsymbol{\Phi}_*,N_\mathrm{bw}(r)\right],
\eea
which we further study in the next section. Combined with \Eq{eq:Power:Spectrum:Exact:FPT}, it gives rise to 
\bea
\label{eq:Power:Spectrum:Pbw}
\calP_\zeta(k) =  \int \dd{\boldmathsymbol{\Phi}}_* \left.
\frac{ \partial P_\mathrm{bw}\left(\boldmathsymbol{\Phi}_*,N_\mathrm{bw}\right)}{\partial N_{\mathrm{bw}}}\right\vert_{N_{\mathrm{bw}}=-\ln(k/k_\uend)}
\left\langle \delta \mathcal{N}^2\right\rangle\left(\boldmathsymbol{\Phi}_*\right)\, .
\eea
At this stage, it is worth noting that at leading order in the low-diffusion limit, the backward probability is simply a Dirac distribution centred on the classical trajectory, $P_\mathrm{bw}(\boldmathsymbol{\Phi}_*,N_\mathrm{bw})=\delta[\boldmathsymbol{\Phi}_*-\boldmathsymbol{\Phi}_\ucl(k)]$, where the function $\boldmathsymbol{\Phi}_\ucl(k)$ was introduced in \Eq{eq:classical:link:k:Phi}. By plugging this expression into \Eq{eq:Power:Spectrum:Pbw}, after integration by parts, one recovers \Eq{eq:PowerSpectrum:Low:Diffusion:Limit}, which is a good consistency check.
\subsection{Backward probability}
\label{sec:Pbw}
In this section, we explain how the backward probability can be computed from the solutions of the Fokker-Planck and adjoint Fokker-Planck equations studied in \Sec{sec:stoch_deltaN_one}. For notational convenience, we use $\mathbb {P}(\cdot)$ to denote the probability (density) associated to the event written in the argument in general. For instance, we have $\mathbb{P}[\boldmathsymbol{\Phi}(N)=\widetilde{\boldmathsymbol{\Phi}}\vert \boldmathsymbol{\Phi}(N_\uin)=\boldmathsymbol{\Phi}_\uin]=P(\widetilde{\boldmathsymbol{\Phi}},N\vert \boldmathsymbol{\Phi}_\uin,N_\uin)$, which was introduced in \Eq{eq:Fokker:Planck}, and $\mathbb {P}[\mathcal{N}(\boldmathsymbol{\Phi})=\tilde{N}]=\Pfpt(\tilde{N},\boldmathsymbol{\Phi})$, which was introduced in \Eq{eq:adjoint:FP}. 

Let us consider the subset of Langevin realisations originating from the original patch at $\boldmathsymbol{\Phi}_0$ that realise at least $N_{\mathrm{bw}}$ inflationary \efolds. The backward probability $P_\mathrm{bw}(\boldmathsymbol{\Phi}_*,N_\mathrm{bw})$ corresponds to the fraction of those realisations for which the value of the fields $N_{\mathrm{bw}}$ \efolds~before the end of inflation is $\boldmathsymbol{\Phi_*}$, so
\bea
\kern-2em
\label{eq:barwardProba:interm:1}
P_\mathrm{bw}\left(\boldmathsymbol{\Phi}_*,N_\mathrm{bw}\right)=\int_{N_\mathrm{bw}}^\infty \dd N_\utot \mathbb {P}\left[\boldmathsymbol{\Phi}(N_\utot-N_\mathrm{bw})=\boldmathsymbol{\Phi}_*, \mathcal{N}\left(\boldmathsymbol{\Phi}_0\right)=N_\utot \vert  \mathcal{N}\left(\boldmathsymbol{\Phi}_0\right)>N_\mathrm{bw}\right]
\eea
where one averages over the value of $\mathcal{N}(\boldmathsymbol{\Phi}_0)$. Since the integral is performed over $N_\utot>N_\mathrm{bw}$, the condition $\mathcal{N}(\boldmathsymbol{\Phi}_0)=N_\utot$ guarantees that $\mathcal{N}(\boldmathsymbol{\Phi}_0)>N_{\mathrm{bw}}$ so the integrand of \Eq{eq:barwardProba:interm:1} is of the form $\mathbb{P}(A,B\vert C)$ with $B \Rightarrow C$. Using the chains rule, it can thus be written as $\mathbb{P}(A,B\vert C)=\mathbb{P}(A,B,C)/\mathbb{P}(C)=\mathbb{P}(A,B)/\mathbb{P}(C)$. Using the chain rule one more time, it is given by $\mathbb{P}(A,B\vert C)=\mathbb{P}(A\vert B)\mathbb{P}(B)/\mathbb{P}(C)$, hence
\bea
\kern-2em
P_\mathrm{bw}\left(\boldmathsymbol{\Phi}_*,N_\mathrm{bw}\right)=
\dfrac{\displaystyle
\int_{N_\mathrm{bw}}^\infty \dd N_\utot
\mathbb {P}\left[\boldmathsymbol{\Phi}(N_\utot-N_\mathrm{bw})=\boldmathsymbol{\Phi}_*  \vert \mathcal{N}\left(\boldmathsymbol{\Phi}_0\right)=N_\utot  \right]\Pfpt\left(N_\utot,\boldmathsymbol{\Phi}_0\right)}{\displaystyle
\int_{N_\mathrm{bw}}^\infty \dd N_\utot\Pfpt\left(N_\utot,\boldmathsymbol{\Phi}_0\right)
}\, .
\eea
The first term in the integrand of the numerator is a probability associated to a past event under a future condition, which makes it difficult to apprehend. Instead, it is simpler to use Baye's theorem and rewrite it in terms of the probability of a future event under a past condition,
\bea
\mathbb {P}\left[\boldmathsymbol{\Phi}(N_\utot-N_\mathrm{bw})= \boldmathsymbol{\Phi}_*  \vert \mathcal{N}\left(\boldmathsymbol{\Phi}_0\right)=N_\utot  \right] & = &
\mathbb {P}\left[\mathcal{N}\left(\boldmathsymbol{\Phi}_0\right)=N_\utot \vert \boldmathsymbol{\Phi}(N_\utot-N_\mathrm{bw})=\boldmathsymbol{\Phi}_*    \right]
\nonumber \\   & & \times 
 \frac{\mathbb {P}\left[\boldmathsymbol{\Phi}(N_\utot-N_\mathrm{bw})=\boldmathsymbol{\Phi}_*  \right]}{\Pfpt\left(N_\utot,\boldmathsymbol{\Phi}_0\right)}\, .
\eea
Because of the Markovian property of the process we consider, the probability appearing on the right-hand side in the first line can be written as $\mathbb {P}[\mathcal{N}(\boldmathsymbol{\Phi}_0)=N_\utot \vert \boldmathsymbol{\Phi}(N_\utot-N_\mathrm{bw})=\boldmathsymbol{\Phi}_*  ]=\mathbb {P}[\mathcal{N}(\boldmathsymbol{\Phi}_*)=N_\mathrm{bw}]=\Pfpt(N_\mathrm{bw},\boldmathsymbol{\Phi}_*)$, and the probability appearing in the numerator in the second line is nothing but $\mathbb {P}[\boldmathsymbol{\Phi}(N_\utot-N_\mathrm{bw})=\boldmathsymbol{\Phi}_*  ]=P(\boldmathsymbol{\Phi}_* ,N_\utot-N_{\mathrm{bw}}\vert \boldmathsymbol{\Phi}_0,0)$, where we set the time on the original patch to zero without loss of generality. Combining the above results, one obtains
\bea
\label{Pbw_revised}
P_\mathrm{bw}\left(\boldmathsymbol{\Phi}_*,N_\mathrm{bw}\right)=
\Pfpt\left(N_\mathrm{bw},\boldmathsymbol{\Phi}_*\right)
\frac{\displaystyle\int_{0}^\infty \dd N P(\boldmathsymbol{\Phi}_* ,N\vert \boldmathsymbol{\Phi}_0,0)}{\displaystyle\int_{N_\mathrm{bw}}^\infty \dd N_\utot\Pfpt\left(N_\utot,\boldmathsymbol{\Phi}_0\right)}\, ,
\eea 
where in the numerator, we have performed the change of integration variable $N=N_\utot-N_{\mathrm{bw}}$. Several remarks are in order regarding this expression. First, one can see that as announced at the beginning of this section, it provides a formula to compute the backwards probability from the knowledge of $P$ and $\Pfpt$ only, \ie from the solutions of the Fokker-Planck and adjoint Fokker-Planck equations. Second, one can check that it is normalised to unity, \ie $\int \dd\boldmathsymbol{\Phi}_* P_\mathrm{bw}(\boldmathsymbol{\Phi}_*,N_\mathrm{bw}) =1$. Third, the term in the denominator corresponds to the probability that at least $N_{\mathrm{bw}}$ \efolds~are realised starting from $\boldmathsymbol{\Phi}_0$. In inflationary models where an arbitrarily large number of \efolds~can be realised (as in large-field or plateau potentials, but not in hilltop potentials), by setting $\boldmathsymbol{\Phi}_0$ high enough in the potential, this term can therefore be brought to values arbitrarily close to one, in which case it can be discarded. Fourth, for one-dimensional setups, the term in the numerator does not depend on $\boldmathsymbol{\Phi}_0$ because of the Markovian nature of the process.\footnote{
To demonstrate this explicitly, let us consider two values for the initial field value, $\phi_{0A}$ and $\phi_{0B}$, with $\phi_\uend < \phi_{0A}<\phi_{0B}$ in a one-dimensional setup. The field distribution initiated from $\phi_{0B}$, $P(\phi,N\vert\phi_{0B},0)$, can be expressed in terms of that from $\phi_{0A}$, $P(\phi,N\vert\phi_{0A},0)$, as $P(\phi,N\vert\phi_{0B},0) = \int \dd\calN\, P(\phi,N-\calN\vert\phi_{0A},0)\theta(N - \calN)\Pfpt(\calN)$ for $\phi < \phi_{0A}$, where $\Pfpt(\calN)$ stands for the first passage time distribution from $\phi_{0B}$ to $\phi_{0A}$. This relation relies on the Markovian property and the fact that the system is one dimensional. By integrating both sides over $N$, one obtains $\int_0^\infty \dd N \,P(\phi,N\vert\phi_{0B},0) = \int_0^\infty \dd N\, P(\phi,N\vert\phi_{0A},0)$.\label{footnote:phi0:single_field}}
This confirms that observable quantities depend on initial conditions for multiple-field systems only. 
\section{Imprint of small-scale diffusion on the large-scale power spectrum}
\label{sec:SmallScales:Impact:LargeScales}
In this section, we apply the results derived in \Sec{sec:PS} to the case where the scales at which the power spectrum is observed cross out the Hubble radius during inflation at a time when quantum diffusion plays a negligible role. In other words, we assume that the backward probability $P_\mathrm{bw}[\boldmathsymbol{\Phi}_*,N_\mathrm{bw}(k)]$ takes most of its support at values of $\boldmathsymbol{\Phi}_*$ where the potential gradient is the main driver of the fields dynamics. This is the case for the scales observed in the CMB in most inflationary models. However, if quantum diffusion plays an important role closer to the end of inflation, which occurs \eg in models producing primordial black holes, the backward probability may be widely spread, and as argued in \Sec{sec:Stoch:DeltaN:review}, the standard formula does not apply in that case. We therefore want to investigate how quantum diffusion at small scales distort the power spectrum at large scales in such scenarios.
\subsection{Averaging the classical power spectrum}
Let us first note that \Eq{Pbw_revised} can be plugged into \Eq{eq:Power:Spectrum:Pbw} to derive a compact formula for the power spectrum. In the limit where $\boldmathsymbol{\Phi}_0$ is taken sufficiently high in the potential and the denominator of \Eq{Pbw_revised} can be discarded, the derivation with respect to $N_{\mathrm{bw}}$ appearing in \Eq{eq:Power:Spectrum:Pbw} only acts on the term $\Pfpt(N_{\mathrm{bw}},\boldmathsymbol{\Phi}_*)$ of \Eq{Pbw_revised}, and according to \Eq{eq:adjoint:FP}, it gives rise to $\mathcal{L}_{\mathrm{FP}}^\dagger(\boldmathsymbol{\Phi}_*)\cdot\Pfpt(N_{\mathrm{bw}},\boldmathsymbol{\Phi}_*) $. By using the definition of the adjoint Fokker-Planck operator in terms of the Fokker-Planck operator, see below \Eq{eq:adjoint:FP}, one obtains
\bea
\label{eq:Power:Spectrum:Pbw:large:Phi0}
\calP_{\zeta}(k)=\int \dd\boldmathsymbol{\Phi_*} \Pfpt\left[N_{\mathrm{bw}}(k),\boldmathsymbol{\Phi_*}\right]
\int_0^\infty \dd N
\mathcal{L}_\mathrm{FP}\left(\boldmathsymbol{\Phi_*}\right)\cdot \left[ \left\langle \delta\mathcal{N}^2\right\rangle \left(\boldmathsymbol{\Phi_*}\right)P\left(\boldmathsymbol{\Phi}_* ,N\vert \boldmathsymbol{\Phi}_0,0\right)\right]\, .
\eea
In this expression, the Fokker-Planck operator, which we recall is a second-order partial differential operator, acts on the product of two terms, namely $\langle\delta\mathcal{N}^2\rangle (\boldmathsymbol{\Phi_*})$ and $P(\boldmathsymbol{\Phi}_* ,N\vert \boldmathsymbol{\Phi}_0,0)$. When acting directly on $P(\boldmathsymbol{\Phi}_* ,N\vert \boldmathsymbol{\Phi}_0,0)$, because of \Eq{eq:Fokker:Planck}, it returns $(\partial /\partial N) P(\boldmathsymbol{\Phi}_* ,N\vert \boldmathsymbol{\Phi}_0,0)$, for which the integral over $N$ can be readily performed in \Eq{eq:Power:Spectrum:Pbw:large:Phi0} and provides a vanishing contribution.\footnote{The lower bound is $P(\boldmathsymbol{\Phi}_*,0\vert\boldmathsymbol{\Phi}_0,0)=\delta(\boldmathsymbol{\Phi}_*-\boldmathsymbol{\Phi}_0)$, so the term $\Pfpt[N_{\mathrm{bw}}(k),\boldmathsymbol{\Phi_*}]$ has to be evaluated at $\boldmathsymbol{\Phi}_*=\boldmathsymbol{\Phi}_0$ where it vanishes for finite $N_{\mathrm{bw}}$ since $\boldmathsymbol{\Phi}_0$ has been set asymptotically high up in the potential. The upper bound $P(\boldmathsymbol{\Phi}_*,\infty\vert\boldmathsymbol{\Phi}_0,0)$ vanishes since all realisations exit the inflating domain after a finite number of \efolds.} This allows one to remove some of the terms appearing in \Eq{eq:Power:Spectrum:Pbw:large:Phi0}. In the regime of low diffusion, the second-order terms (\ie those involving second derivatives with respect to the field values) can be neglected as they account for quantum diffusion, so only the drift term of the Fokker-Planck equation remains, and one finds
\bea
\label{eq:Power:Spectrum:Pbw:large:Phi0:LowDiffusion:interm}
\calP_\zeta(k)\simeq -\int\dd\boldmathsymbol{\Phi}_* P_\mathrm{bw}\left[\boldmathsymbol{\Phi}_*,N_{\mathrm{bw}}(k)\right]
\boldmathsymbol{F}_\ucl(\boldmathsymbol{\Phi}_*)\cdot \frac{\partial}{\partial\boldmathsymbol{\Phi}_*}\left\langle \delta\mathcal{N}^2\right\rangle \left(\boldmathsymbol{\Phi_*}\right)
\eea
where we recall that the drift function $\boldmathsymbol{F}_\ucl$ was introduced in \Eq{eq:Langevin}, and where the dot stands for a scalar product, namely $\boldmathsymbol{F}_\ucl\cdot (\partial/\partial \boldmathsymbol{\Phi}_*)=\sum_i F_{\ucl,i}(\partial/\partial\Phi_i)$ where one sums over all the fields contained in the vector $\boldmathsymbol{\Phi}$. Since $\boldmathsymbol{F}_\ucl$ is, by definition, $\partial\boldmathsymbol{\Phi}/\partial N$ along the classical trajectory, one can rewrite the derivative with respect to $\boldmathsymbol{\Phi}_*$ in \Eq{eq:Power:Spectrum:Pbw:large:Phi0:LowDiffusion:interm} as a derivative with respect to $N$, leading to
\bea
\label{eq:Power:Spectrum:Pbw:large:Phi0:LowDiffusion}
\calP_\zeta(k)\simeq\int\dd\boldmathsymbol{\Phi}_* P_\mathrm{bw}\left[\boldmathsymbol{\Phi}_*,N_{\mathrm{bw}}(k)\right]
\calP_{\zeta,\ucl}(\boldmathsymbol{\Phi}_*).
\eea
In this expression, the classical power spectrum $\calP_{\zeta,\ucl}$ corresponds to \Eq{eq:PowerSpectrum:Low:Diffusion:Limit}, obtained in the low-diffusion limit. The physical interpretation of \Eq{eq:Power:Spectrum:Pbw:large:Phi0:LowDiffusion} is rather clear: it corresponds to the standard formula for the power spectrum evaluated at $\boldmathsymbol{\Phi}_*$, and averaged over all possible values of $\boldmathsymbol{\Phi}_*$ reached $N_{\mathrm{bw}}(k)$ \efolds~before the end of inflation. Let us however stress that, although intuitive, this formula only holds when quantum diffusion is low at the scale one considers: in general, the power spectrum does \emph{not} result from an averaging procedure of the type of \Eq{eq:Power:Spectrum:Pbw:large:Phi0:LowDiffusion}, and one should rather use \Eq{eq:Power:Spectrum:Pbw:large:Phi0}, which features a more involved structure.
\subsection{A quantum well between two classical regions}
\label{sec:QuantumWell:Sandwiched}
\begin{figure}[t]
\begin{center}
\includegraphics[width=0.8\textwidth]{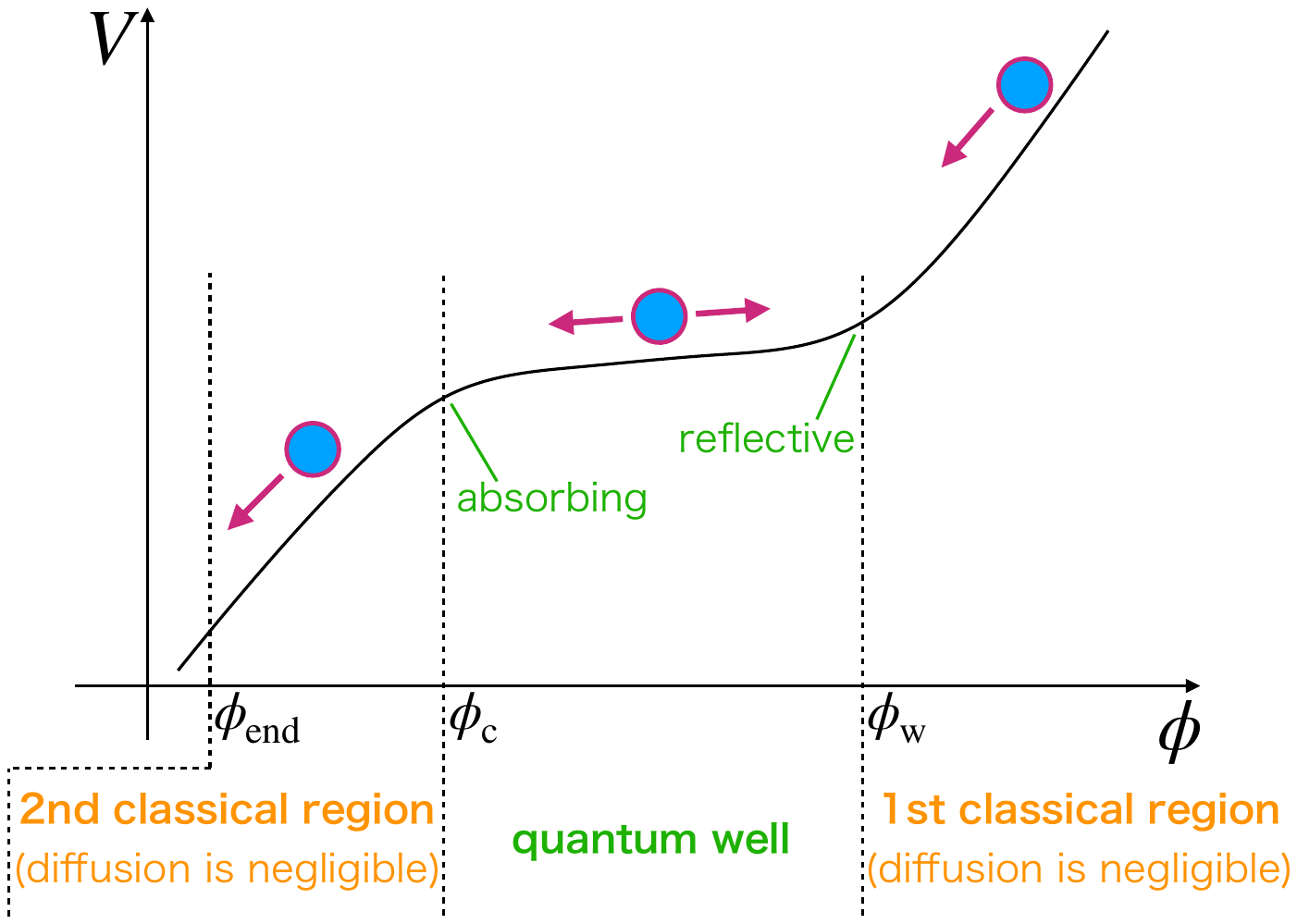}
\caption{Sketch of the single-field potential studied in \Sec{sec:QuantumWell:Sandwiched}, where quantum diffusion is only effective inside a ``quantum well'', that is surrounded by two regions where the field is only driven by the potential gradient.
}  
\label{fig:potential_well}
\end{center}
\end{figure}
For explicitness, let us restrict the analysis to one-field slow-roll setups, for which the Langevin equation~\eqref{eq:Langevin} reads
\bea
\label{eq:Langevin:SingleField:SlowRoll}
\frac{\dd \phi}{\dd N} = -\Mp^2 \frac{v'}{v}+\frac{H}{2\pi}\xi(N)
\eea
where $\xi$ is a white Gaussian noise with vanishing mean and unit variance. The potential we consider is depicted in \Fig{fig:potential_well}, where quantum diffusion is only effective inside the region comprised between $\phi_\uc$ and $\phi_\uw$ that we call the ``quantum well''. This quantum well is surrounded by two other regions where we assume that the stochastic noise plays a negligible role compared to the potential gradient and can therefore be neglected. This corresponds for instance to models with a flat inflection point close to the end of inflation~\cite{Garcia-Bellido:2017mdw, Germani:2017bcs}, where large perturbations (that possibly later collapse into primordial black holes) are produced within the well, while the CMB scales emerge in the first classical region, at $\phi>\phi_\uw$.

Let us introduce a few notations. For $\phi>\phi_\uw$, let $N_{\ucl,1}(\phi)$ be the number of \efolds~that is classically realised from $\phi$ to $\phi_\uw$, and similarly, for $\phi<\phi_\uc$, let $N_{\ucl,2}(\phi)$ denote the number of \efolds~that is classically realised from $\phi$ to $\phi_\uend$. In practice, they are given by \Eq{eq:meanN:class}, where, for $N_{\ucl,1}(\phi)$, the lower bound of the integral has to be set to $\phi_\uw$. These functions are inverted as $\phi_{\ucl,1}(N)$ and $\phi_{\ucl,2}(N)$, which return the field value $N$ \efolds~before reaching $\phi_\uw$ and $\phi_\uend$ respectively.
 In addition, for $\phi_\uc<\phi<\phi_\uw$, the distribution associated to the time of first passage through $\phi_\uc$ is noted $\Pfpt^\mathrm{well}(\mathcal{N},\phi)$. Let us note that since the field can only decrease at $\phi<\phi_\uc$, it can never return into the well once it has escaped from it, and likewise, since the field can only decrease at $\phi>\phi_\uw$, it can never exit the well from above. In practice, this can be modelled by setting an absorbing boundary at $\phi_\uc$ and a reflective boundary at $\phi_\uw$. Finally, the probability associated to the field value after it has spent $N$ \efolds~inside the well is noted $P^\uwell(\phi,N)= P(\phi,N\vert \phi_\uw,0)$. 

In \Sec{sec:Flat:Well}, the calculation will be specified to the case where the quantum well is exactly flat, but the present considerations are still generic, the only assumption is that the stochastic noise can be neglected outside the well.

%%%%%%%%
\subsubsection*{Backward probability}
\label{subsec:quantum_classical_Pbw}
%%%%%%%%
Let us first evaluate the backward probability in this model. We assume that $\phi_0$ is set sufficiently high up in the potential such that the denominator of \Eq{Pbw_revised} can be discarded. In practice, this is guaranteed if one considers scales such that $N_\ubw(k)\leq N_{\ucl, 2}(\phi_\uc) + N_{\ucl, 1}(\phi_0)$. Since field space is one dimensional, we therefore expect the result not to depend on $\phi_0$, see footnote~\ref{footnote:phi0:single_field}.

If $N_\ubw \leq N_{\ucl,2}(\phi_\uc)$, the field is necessarily in the second deterministic region $N_\ubw$ \efolds~before the end of inflation, and the backward probability is simply
\bea
\label{eq:Pbw:deterministic}
	P_\ubw(\phi_*, N_\ubw) = \delta\left[
	\phi_* - \phi_{\ucl, 2}(N_\ubw)
	\right]
	\quad\mathrm{for}\quad
	\phi_\uend \leq \phi_*  \leq  \phi_\uc
	\,.
\eea

If $N_\ubw \geq N_{\ucl,2}(\phi_\uc)$, the field is either in the quantum well or in the first deterministic region $N_\ubw$ \efolds~before the end of inflation.
The solution of the Fokker-Planck equation, for $\phi>\phi_\uc$, is given by
\bea
\label{FP_sol_quantum_classical}
	P(\phi, N\vert \phi_0,0) =
	\begin{cases}
		P^\uwell\left[\phi, N - N_\mathrm{cl, 1}\left(\phi_0\right)\right] \theta \left[N - N_\mathrm{cl, 1}(\phi_0)\right]
		& \mathrm{for} \quad
		\phi_\uc \leq \phi \leq \phi_\mathrm{w} \\
		\delta\left\lbrace \phi - \phi_\mathrm{cl 1}\left[N_{\ucl, 1}(\phi_0) - N\right]\right\rbrace \theta \left[N_\mathrm{cl, 1}(\phi_0) - N\right]
		& \mathrm{for} \quad
		\phi \geq \phi_\mathrm{w}
	\end{cases}
	\, ,
\eea
where $\theta$ is the Heaviside function, and we do not give the solution for  $\phi<\phi_\uc$ since we will not use it. In order to evaluate the backward probability with \Eq{Pbw_revised}, on the one hand, one needs to integrate this probability over $N$, giving rise to
\bea
\label{int_FP_sol_quantum_classecal}
	\int_0^\infty \dd N\,P(\phi_*, N\vert \phi_0,0) =
	\begin{cases}
		\int_0^\infty \dd N\, P^\uwell(\phi_*, N )
		&\quad \mathrm{for} \quad
		\phi_\uc \leq \phi_* \leq \phi_\mathrm{w} \\
		N_{\ucl,1}'\left(\phi_*\right)
		&\quad \mathrm{for} \quad
		\phi_* > \phi_\mathrm{w}
	\end{cases}
	\,.
\eea
On the other hand, \Eq{Pbw_revised} involves the probability density function of the first passage time ${\Pfpt} (N, \phi_*)$, which is given by
\bea
\label{Pfpt_quantum_classical}
\hspace{-10mm}
	&&{\Pfpt} (N, \phi_*) = \nonumber\\
\hspace{-10mm}
	&&\begin{cases}
		\Pfpt^\uwell \left[N-N_{\ucl, 2}(\phi_\uc), \phi_*\right]
		& \mathrm{for} \quad
		\phi_\uc \leq \phi_* \leq \phi_\mathrm{w} \\
		\Pfpt^\uwell \left[N - N_\mathrm{cl, 1}(\phi_*) - N_\mathrm{cl, 2}(\phi_\uc), \phi_\mathrm{w}\right]\theta \left[N - N_\mathrm{cl, 1}(\phi_*) - N_\mathrm{cl, 2}(\phi_\uc)\right]
		& \mathrm{for} \quad
		\phi_* \geq \phi_\mathrm{w}
	\end{cases}
	.
\eea
By plugging  \Eqs{int_FP_sol_quantum_classecal}  and \eqref{Pfpt_quantum_classical} into \Eq{Pbw_revised}, it can thus be computed for $N_\ubw \geq N_{\ucl,2}(\phi_\uc)$, and combining the result with \Eq{eq:Pbw:deterministic}, one obtains
\bea
\label{Pbw_quantum_classical}
	&&P_\mathrm{bw}\prn{\phi_*, N_\mathrm{bw}} = \nonumber\\
	&&
	\begin{cases}
		\delta\left[
		\phi_* - \phi_{\ucl, 2}(N_\ubw)
		\right] 
		\theta\left[
		N_{\ucl,2}(\phi_\uc) - N_\ubw
		\right]
	&\mathrm{for}\quad
	\phi_\uend \leq \phi_*  \leq  \phi_\uc \\
		\Pfpt^\uwell \left[N_\ubw-N_{\ucl, 2}(\phi_\uc), \phi_*\right]
		\int_0^\infty \dd N\, P^\uwell(\phi_*, N )
		\theta \left[N_\ubw - N_\mathrm{cl, 2}(\phi_\uc)\right]
		&\mathrm{for} \quad
		\phi_\uc \leq \phi_* \leq \phi_\mathrm{w} \\
		N_{\ucl,1}'\left(\phi_*\right)
		\Pfpt^\uwell \left[N_\ubw - N_\mathrm{cl, 1+2}(\phi_*), \phi_\mathrm{w}\right]
		\theta \left[N_\ubw - N_\mathrm{cl, 1+2}(\phi_*)\right]
		&\mathrm{for} \quad
		\phi_* \geq \phi_\mathrm{w}
	\end{cases}
	\, ,
	\nonumber \\
\eea
where we have  defined $N_\mathrm{cl, 1+2}(\phi_*) \equiv N_\mathrm{cl, 1}(\phi_*) + N_\mathrm{cl, 2}(\phi_\uc)$.
%%%%%%%%%%%%%%%%%%%%%%%%%%%%%%%%%%%%%%%%%%%%%%%%%%%%%
\subsubsection*{Power spectrum}
\label{subsec:quantum_classical_PS}
%%%%%%%%%%%%%%%%%%%%%%%%%%%%%%%%%%%%%%%%%%%%%%%%%%%%%
In order to evaluate the power spectrum~\eqref{eq:Power:Spectrum:Pbw}, one also needs to compute $\langle\delta\mathcal{N}^2\rangle(\phi_*)$. As explained above, the boundaries placed at $\phi_\uc$ and $\phi_\uw$ are one-way boundaries (the field can cross them once only and only from above), so the numbers of \efolds~realised in each of the three regions are independent random variables (invoking again the Markovian nature of the process). As a consequence, one has
\bea
\kern-2em
\label{deltaN2_quantum_classical}
\left\langle\delta\mathcal{N}^2\right\rangle\left(\phi_*\right)=
\begin{cases}
\left\langle \delta N_{\ucl,2}^2 \right\rangle\left(\phi_*\right)\quad & \text{for} \quad \phi_*<\phi_\uc\\
\left\langle \delta N_{\ucl,2}^2 \right\rangle\left(\phi_\uc\right)+\left\langle \delta \mathcal{N}_{\uwell}^2 \right\rangle\left(\phi_*\right)\quad & \text{for} \quad \phi_\uc<\phi_*<\phi_\uw\\
\left\langle \delta N_{\ucl,2}^2 \right\rangle\left(\phi_\uc\right)+\left\langle \delta \mathcal{N}_{\uwell}^2 \right\rangle\left(\phi_\uw\right)+\left\langle \delta N_{\ucl,1}^2 \right\rangle\left(\phi_*\right)\quad & \text{for} \quad \phi_*>\phi_\uw
\end{cases}\, ,
\eea 
with $\langle \delta N_{\ucl,1}^2 \rangle$ and $\langle \delta N_{\ucl,2}^2 \rangle$ given by \Eq{eq:meandeltaN2:class}, where, in the case of $\langle \delta N_{\ucl,1}^2 \rangle$, the lower bound of the integral needs to be replaced with $\phi_\uc$.

If the scale $k$ is such that $N_\ubw(k)<N_{\ucl,2}(\phi_\uc)$, only the first branch ($\phi_*\leq \phi_\uc$) of \Eq{Pbw_quantum_classical} contributes to the integral of \Eq{eq:Power:Spectrum:Pbw}. By acting $\partial/\partial N_\ubw$ onto $P_\ubw$ given in \Eq{Pbw_quantum_classical}, one obtains $-\delta'[\phi_*-\phi_{\ucl,2}(N_\ubw)]/N_{\ucl,2}'(\phi_*)$. Plugging the result into \Eq{eq:Power:Spectrum:Pbw}, and making use of \Eq{deltaN2_quantum_classical}, after integration by parts one obtains $\calP_\zeta(k)=\langle \delta N_{\ucl,2}^2  \rangle'[\phi_{\ucl,2}(N_\ubw)]/N_{\ucl,2}'[\phi_{\ucl,2}(N_\ubw)]$. Using \Eqs{eq:meanN:class} and~\eqref{eq:meandeltaN2:class}, this gives rise to \Eq{eq:Pzeta:classical:SingleField:SlowRoll}, so one recovers the standard formula for the power spectrum in the low-diffusion limit, which is a good consistency check.

If the scale $k$ is such that $N_\ubw(k)>N_{\ucl,2}(\phi_\uc)$, the second ($\phi_\uc<\phi_*<\phi_\uw$) and third ($\phi>\phi_\uw$) branches of \Eq{Pbw_quantum_classical} contribute to the integral of \Eq{eq:Power:Spectrum:Pbw}. We denote their contributions by $\calP_\zeta^{(2)}$ and $\calP_\zeta^{(3)}$ respectively. We cannot say much about $\calP_\zeta^{(2)}$ without specifying the shape of the inflationary potential inside the quantum well (this will be done in \Sec{sec:Flat:Well}) so let us focus on $\calP_\zeta^{(3)}$. When acting $\partial/\partial N_\ubw$ on the third branch of \Eq{Pbw_quantum_classical}, one obtains two terms, one where $\partial/\partial N_\ubw$ acts on $\Pfpt^\uwell$ and one where $\partial/\partial N_\ubw$ acts on the Heaviside function. The second term is proportional to $\delta[N_\ubw-N_{\ucl,1+2}(\phi_*)]$ so it involves $\Pfpt^\uwell (0,\phi_\uw)$, which vanishes (the probability to cross the entire well in no time at all is necessarily zero). Only remains the first term, and by plugging the result into \Eq{eq:Power:Spectrum:Pbw}, one obtains
\bea
\kern-2em
\calP_\zeta^{(3)} (k)=  \int_{\phi_\uw}^{\phi_{\ucl,1}[N_\ubw-N_{\ucl,2}(\phi_\uc)]} \dd\phi_* N_{\ucl,1}'(\phi_*) \frac{\partial\Pfpt^\uwell\left[N_\ubw - N_\mathrm{cl, 1+2}(\phi_*), \phi_\mathrm{w}\right] }{\partial N_\ubw} \left\langle \delta\mathcal{N}^2\right\rangle (\phi_*).
\eea
Let us then perform a change of integration variable and label the field value $\phi_*$ via the corresponding classical \efold~number, $N_\ucl=N_{\ucl,1+2}(\phi_*)$. This allows us to rewrite the above expression as
\bea
\calP_\zeta^{(3)} (k)=  \int_{N_{\ucl,2}(\phi_\uc)}^{N_\ubw} \dd N_\ucl \frac{\partial\Pfpt^\uwell\left(N_\ubw - N_\ucl, \phi_\mathrm{w}\right) }{\partial N_\ubw} \left\langle \delta\mathcal{N}^2\right\rangle \left\lbrace \phi_{\ucl,1}\left[N_\ucl-N_{\ucl,2}(\phi_\uc)\right]\right\rbrace.
\eea
An important remark is that since $\Pfpt^\uwell$ depends on $N_\ubw$ only through the combination $N_\ubw-N_\ucl$, acting $\partial/\partial N_\ubw$ on it is equivalent to acting $-\partial/\partial N_\ucl$. This allows one to integrate by parts, leading to
\bea
\calP_\zeta^{(3)}(k)&=&\Pfpt^\uwell\left[N_\ubw-N_{\ucl,2}(\phi_\uc),\phi_\uw \right]\left\langle \delta\mathcal{N}^2\right\rangle\left(\phi_\uw\right)
\nonumber \\ & & \kern-3em
+\int_{N_{\ucl,2}(\phi_\uc)}^{N_\ubw} \dd N_\ucl \Pfpt^\uwell\left(N_\ubw - N_\ucl, \phi_\mathrm{w}\right) \phi_{\ucl,1}'\left[N_\ucl-N_{\ucl,2}(\phi_\uc)\right] \left\langle \delta\mathcal{N}^2\right\rangle' \left\lbrace \phi_{\ucl,1}\left[N_\ucl-N_{\ucl,2}(\phi_\uc)\right]\right\rbrace
\nonumber \\ 
\eea
where we have used again that $\Pfpt^\uwell (0,\phi_\uw)=0$. In this expression, making use of \Eq{deltaN2_quantum_classical}, one has $\langle \delta\mathcal{N}^2\rangle' =\langle \delta N_{\ucl,1}^2 \rangle'$, and performing the inverse change of integration $N_\ucl \to \phi_*= \phi_{\ucl,1}[N_\ucl-N_{\ucl,2}(\phi_\uc)]$, one obtains
\bea
\label{eqL:calPzeta:3:final}
\calP_\zeta^{(3)}(k)&=&\Pfpt^\uwell\left[N_\ubw-N_{\ucl,2}(\phi_\uc),\phi_\uw \right]\left\langle \delta\mathcal{N}^2\right\rangle\left(\phi_\uw\right)
\nonumber \\ & & \kern-3em
+\int_{\phi_\uw}^{\phi_{\ucl,1}[N_\ubw-N_{\ucl,2}(\phi_\uc)]} \dd \phi_* \Pfpt^\uwell\left[N_\ubw - N_{\ucl,1+2}(\phi_*), \phi_\mathrm{w}\right] 
 \left\langle \delta N_{\ucl,1}^2\right\rangle'  (\phi_*)\, . 
\eea
Let us note that, looking at the third branch of \Eq{Pbw_quantum_classical}, the second term in the above expression can be written as $\int \dd \phi_* P_\ubw(\phi_*,N_\ubw)  \calP_{\zeta,\ucl}(\phi_*)$ where the classical power spectrum $\calP_{\zeta,\ucl}=\langle \delta N_\ucl^2\rangle'/N_\ucl'$ is given in \Eq{eq:Pzeta:classical:SingleField:SlowRoll}. One therefore recovers the standard power spectrum averaged with the backwards probability, see \Eq{eq:Power:Spectrum:Pbw:large:Phi0:LowDiffusion}, but let us stress that this is not the only contribution: there is also the first integrated term of \Eq{eqL:calPzeta:3:final}, and $\calP_\zeta^{(2)}$, which we have not computed yet and to which we now turn our attention. These additional contributions correspond to realisations where, $N_\ubw(k)$ \efolds~before the end of inflation, the field is found inside the quantum well.
\subsection{Case of a flat quantum well}
\label{sec:Flat:Well}
In this section, in order to derive explicit results, we assume for simplicity that the quantum well is exactly flat. One may be concerned that, if the potential is exactly flat, inflation proceeds in the ultra-slow-roll regime and the slow-roll approximation may not be valid. This is however not the case if the approach to the flat region is sufficiently smooth, and in practice, as shown in \Refs{Pattison:2017mbe, Ezquiaga:2019ftu}, a flat well provides a good approximation to several potentials featuring flat or quasi-flat points, for which the width of the equivalent flat well has to be set by the criterion $\left\vert v''  \right\vert v^2\sim {v'}^2$ mentioned above. This toy model is therefore of practical interest. 

By rescaling the field variable as $x=(\phi-\phi_\uc)/(\phi_\uw-\phi_\uc)$ inside the well, where $x$ varies between $0$ and $1$, the Langevin equation~\eqref{eq:Langevin:SingleField:SlowRoll} takes the simple form 
\bea
\label{eq:Langevin:FlatPot}
\frac{\dd x}{\dd N} = \frac{\sqrt{2}}{\mu}\xi(N),\quad\text{where}\quad \mu^2=\frac{\left(\phi_\uw-\phi_\uc\right)^2}{\Mp^2 v_\uwell}
\eea
is the ratio between the squared width of the potential well and its height, in Planckian units. In \App{sec:FP:sol:Flat}, we show that the solution to the Fokker-Planck and adjoint Fokker-Planck problems are given by
\bea
\label{eq:Pwell:final}
P^\uwell(x,N \vert x_\uin,0)&=& \frac{1}{2}\vartheta_2\left[-\frac{\pi}{2}\left(x-x_\uin\right),\ee^{-\frac{\pi^2 N}{\mu^2}}\right]+\frac{1}{2}\vartheta_2\left[-\frac{\pi}{2}\left(x+x_\uin\right),\ee^{-\frac{\pi^2 N}{\mu^2}}\right]\\
\Pfpt^\uwell(x,N)&=&-\frac{\pi}{2\mu^2}\vartheta_2'\left(\frac{\pi}{2}x,\ee^{-\frac{\pi^2 N}{\mu^2}}\right)
\label{eq:Pfptwell:final}
\eea
where $\vartheta_2$ is the second elliptic theta function,\footnote{The elliptic theta functions are defined by~\cite{2011ConPh..52..497T}
\begin{equation}
\begin{aligned}
\label{eq:elliptic:def}
\vartheta_1\left(z,q\right)&=2\sum_{n=0}^\infty (-1)^n q^{\left(n+\frac{1}{2}\right)^2}\sin\left[\left(2n+1\right)z\right]
 \ ,\quad\quad
\vartheta_2\left(z,q\right)=2\sum_{n=0}^\infty  q^{\left(n+\frac{1}{2}\right)^2}\cos\left[\left(2n+1\right)z\right]\, , \\
\vartheta_3\left(z,q\right)&=1+2\sum_{n=1}^\infty  q^{n^2}\cos\left(2nz\right) 
 \ ,\quad\quad
\vartheta_4\left(z,q\right)=1+2\sum_{n=1}^\infty  (-1)^n q^{n^2}\cos\left(2nz\right) 
\, .
\end{aligned}
\end{equation}
\label{footnote:elliptic:def}}
and $\vartheta_2'$ denotes its derivative with respect to its first argument. Note that the expression for $\Pfpt^\uwell$ was already obtained in \Refa{Pattison:2017mbe}, see Eq.~(4.11) of that reference, but in \App{sec:FP:sol:Flat} it is derived with different techniques (namely the ``method of images'', while in \Refa{Pattison:2017mbe} the result is obtained by solving for the characteristic function). In a flat potential, the integrals appearing in \Eq{eq:moments:SingleField:SlowRoll} can also be performed exactly, and for the two first moments, one obtains
\bea
	\label{eq:meanN:well}
	\left<\mathcal N_\uwell \right>(x) &=&
	\frac{\mu^2}{2} \com{1-(1-x)^2}\, , \\
	\left<\delta\mathcal N_\uwell^2 \right> (x)&=&
	\frac{\mu^4}{6} \com{1-(1-x)^4}\, .
	\label{eq:deltaN2:well}
\eea

\begin{figure}[t]
\begin{center}
\includegraphics[width=0.5\textwidth]{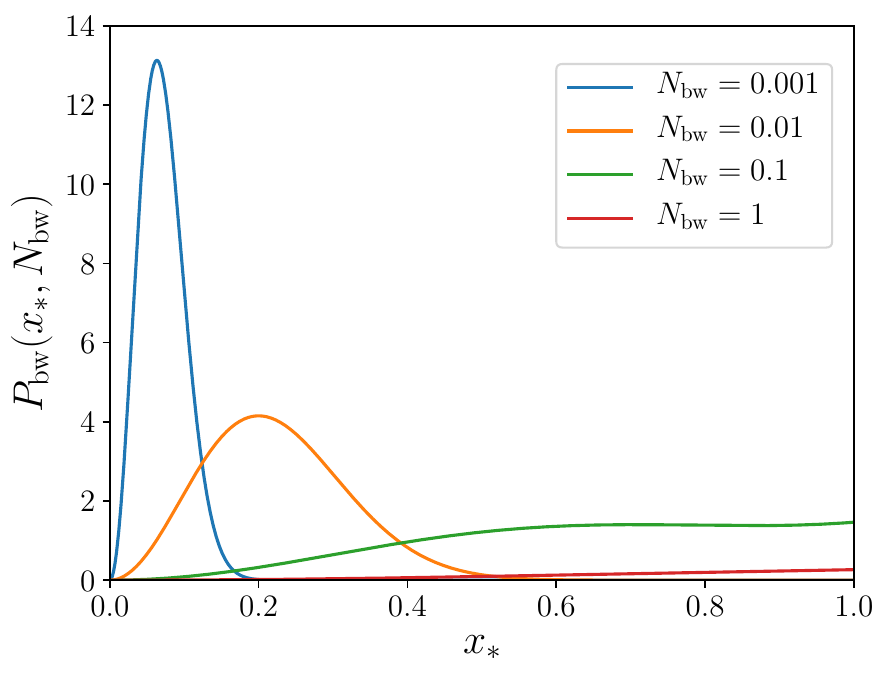}
\caption{Backward probability in a flat potential with an absorbing wall at $x = 0$ and a reflective wall at $x = 1$, computed by means of \Eq{Pbw_reflect_analytic}, with $\mu=1$ and for a few values of $N_\ubw$, assuming that $N_{\ucl,2}(\phi_\uc)=0$ [otherwise the value of $N_\ubw$ is simply shifted by $N_{\ucl,2}(\phi_\uc)$] .}  
\label{fig:Pbw_flat_reflect_mu2}
\end{center}
\end{figure}
These expressions allow one to further specify the second and third branches of the backward probability in \Eq{Pbw_quantum_classical}. The second branch requires to integrate \Eq{eq:Pwell:final} over $N$ when $x_\uin=1$. This is done in details at the end of \App{sec:FP:sol:Flat}, where it is first noted that by setting $x_\uin=1$ in \Eq{eq:Pwell:final}, one obtains $P^\uwell(x,N)=\vartheta_1(\pi x/2,\ee^{-\pi^2 N/\mu^2}) $, which then gives rise to $\int_0^\infty  P^\uwell(x,N) \dd N = \mu^2 x$. When $N_\ubw>N_{\ucl,2}(\phi_\uc)$, the backward probability~\eqref{Pbw_quantum_classical} thus reads
\bea
\label{Pbw_reflect_analytic}
& & P_\ubw\left(\phi_*, N_\ubw\right)= \nonumber \\ & &
\begin{cases}
-\frac{\pi x_*}{2(\phi_\uw-\phi_\uc)}\vartheta_2'\left( \frac{\pi}{2}x_*,\ee^{-\frac{\pi^2}{\mu^2}\left[N_\ubw-N_{\ucl,2}\left(\phi_\uc\right)\right]} \right) &\quad \text{for}\quad \phi_\uc<\phi_*<\phi_\uw\\
- N_{\ucl,1}'\left(\phi_*\right)\frac{\pi}{2\mu^2}\vartheta_2'\left( \frac{\pi}{2},\ee^{-\frac{\pi^2}{\mu^2}\left[N_\ubw-N_{\ucl,1+2}\left(\phi_*\right)\right]} \right)\theta \left[N_\ubw - N_\mathrm{cl, 1+2}(\phi_*)\right]&\quad \text{for}\quad \phi_*>\phi_\uw
\end{cases}\nonumber\\
\eea 
and is displayed in \Fig{fig:Pbw_flat_reflect_mu2} when $\phi_*$ lies within the well (outside the well, the distribution depends on the inflationary potential at $\phi>\phi_\uw$, which we have not specified yet) and for a few values of $N_\ubw$. When $N_\ubw$ is small, the distribution is peaked close to the absorbing boundary, and is in fact similar to the one obtained in \App{subsec:Pbw_revised_flat_abs_wall1} without setting a reflective boundary at $x=1$, see \Fig{fig:Pbw_flat_single_right}. This is because, for small values of $N_\ubw$, the probability to bounce against the reflective wall during the last $N_\ubw$ \efolds~spent in the well is low. Formally, one can indeed show that by expanding \Eq{Pbw_reflect_analytic} in the limit $x_*\ll 1$, one recovers \Eq{Pbw_flat_single_revised_lim}. When $N_\ubw$ increases, the backward distribution becomes more widely spread, and centred over larger field values. Let us also note that by integrating \Eq{Pbw_reflect_analytic} over $\phi_*$, one can check explicitly that it is properly normalised, and that the probability to find the field inside the well $N_\ubw$ \efolds~before the end of inflation is given by
\bea
\label{eq:pwell}
p_\uwell\left(N_\ubw\right)=\frac{4}{\pi}\sum_{n=0}^\infty \frac{(-1)^n}{2n+1}\ee^{-\left(n+\frac{1}{2}\right)^2\frac{\pi^2}{\mu^2}\left[N_\ubw-N_{\ucl,2}(\phi_\uc)\right]}\, .
\eea
When $N_\ubw=N_{\ucl,2}(\phi_\uc)$, this probability equals one as expected, and then it decreases when $N_\ubw$ increases. 

Let us now compute the power spectrum. The contribution coming from the branch $\phi_\uc<\phi_*<\phi_\uw$, denoted $\calP_\zeta^{(2)}$ in \Sec{sec:QuantumWell:Sandwiched}, can be obtained by plugging \Eqs{Pbw_reflect_analytic} and~\eqref{eq:deltaN2:well} into \Eq{eq:Power:Spectrum:Pbw}. It is however more convenient to use \Eq{eq:Power:Spectrum:Pbw:large:Phi0} directly, where the Fokker-Planck operator can be read off from the right-hand side of \Eq{eq:FokkerPlanck:FlatPot:App}, namely $\mathcal{L}_\mathrm{FP}=\partial^2/\partial x^2$. This gives rise to
\bea
\label{eq:calP:2:interm}
\calP_\zeta^{(2)}(k) =& \frac{\pi\mu^2}{3}\int_0^1 \dd x \left(1-x\right)^2\left(5x-2\right)\vartheta_2'\left[\frac{\pi}{2}x,\ee^{-\frac{\pi^2}{\mu^2}N_\ubw(k)}\right]
\nonumber \\ & 
+\frac{\pi\mu^2}{12}\vartheta_2'\left( \frac{\pi}{2},\ee^{-\frac{\pi^2}{\mu^2} \left[N_\ubw(k)-N_{\ucl,2}\left(\phi_\uc\right)\right]}\right)\, .
\eea
The integral in the first line can be performed by expanding the second elliptic function according to its definition~\eqref{eq:elliptic:def}, and by integrating each term individually.
Making use of \Eqs{eq:Pfptwell:final} and\eqref{eq:deltaN2:well}, one can also show that the term in the second line exactly cancels out with the first term in \Eq{eqL:calPzeta:3:final}.
Furthermore, the integral term of $\calP_\zeta^{(3)}$ can be evaluated by plugging \Eqs{eq:Pfptwell:final} and~\eqref{eq:deltaN2:well} into \Eq{eqL:calPzeta:3:final}, and after performing the change of variable $\phi_*\to N_\ubw-N_{\ucl,2}(\phi_\uc)-N_{\ucl,1}(\phi_*)$ in the integral, the sum of $\calP_\zeta^{(2)}$  and $\calP_\zeta^{(3)}$ reads
\bea
\label{eq:Pzeta:FlatWell:final}
\calP_\zeta(k)&=&\frac{4\mu^2}{3}\vartheta_2\left( 0,\ee^{-\frac{\pi^2}{\mu^2} \left[N_\ubw(k)-N_{\ucl,2}\left(\phi_\uc\right)\right]}\right)
\nonumber \\& &
+8\frac{\mu^2}{\pi^3}\sum_{n=0}^\infty \frac{\ee^{-\left(n+\frac{1}{2}\right)^2\frac{\pi^2}{\mu^2} \left[N_\ubw(k)-N_{\ucl,2}\left(\phi_\uc\right)\right]}}{\left(n+\frac{1}{2}\right)^2}\left[5\frac{(-1)^n}{\left(n+\frac{1}{2}\right)}-4\pi \right]
\nonumber \\& &
-\frac{1}{2\pi} \int_{\ee^{-\frac{\pi^2}{\mu^2} \left[N_\ubw(k)-N_{\ucl,2}\left(\phi_\uc\right)\right]}}^1 \frac{\dd q}{q} \vartheta_2'\left(\frac{\pi}{2},q\right) \calP_{\zeta,\ucl}\, .
\eea
%%%
This expression applies for scales $k$ such that $N_\ubw(k)>N_{\ucl,2}(\phi_\uc)$, and when $N_\ubw(k)<N_{\ucl,2}(\phi_\uc)$, we recall that \Eq{eq:Pzeta:classical:SingleField:SlowRoll} applies, see the discussion below \Eq{deltaN2_quantum_classical}.
In the last term, $\calP_{\zeta,\ucl}=\langle \delta N_\ucl^2\rangle'/N_\ucl'$ corresponds to the standard formula for the power spectrum, see \Eq{eq:Pzeta:classical:SingleField:SlowRoll}. It needs to be evaluated at the value $\phi_*$ related to $q$ through the change of variables we just mentioned. 
Note that this last term can also be written as $\int_{\phi_\uw}^{\infty} \dd\phi_* P_\ubw\prn{\phi_*, N_\ubw}\mathcal{P}_{\zeta,\ucl}(\phi_*)$, so it corresponds to the standard power spectrum averaged over the first classical part of the inflationary potential with the backwards probability, see \Eq{eq:Power:Spectrum:Pbw:large:Phi0:LowDiffusion}. This is however not the only contribution, and the power spectrum receives corrections from the quantum well directly, in the form of the terms displayed in the first two lines. One notices that if, $N_\ubw(k)$ \efolds~before the end of inflation, the probability to find the field inside the well is low, \ie if $p_\uwell(N_\ubw)\ll 1$, where $p_\uwell(N_\ubw)$ is given in \Eq{eq:pwell}, then the correction coming from those terms is small too. This happens when the number of \efolds~elapsed before the second classical part is much larger than the mean number of \efolds~spent in the well, given in \Eq{eq:meanN:well}. In the limit where this is the case, \ie when $N_\ubw-N_{\ucl,2}(\phi_\uc)\gg \mu^2$, the terms with $n=0$ and $n=1$ dominate in the infinite sums appearing in those corrective terms, which thus provide a contribution approximated by
\bea
\label{eq:Pzeta:Correction}
\Delta \calP_\zeta(k)\simeq \underbrace{\frac{960 - 384\pi + 8\pi^3}{3\pi^3}}_{\simeq 0.02} \mu^2\ee^{-\frac{\pi^2}{4\mu^2} \left[N_\ubw(k)-N_{\ucl,2}\left(\phi_\uc\right)\right]}\, .
\eea
In practice, the amplitude of the correction coming from the direct contribution of the well can thus be assessed by comparing this value to the amplitude of the power spectrum observed in the CMB, $\calP_\zeta\simeq 2.1\times 10^{-9}$.
\subsubsection*{Quadratic potential}
In order to illustrate our analytical result~\eqref{eq:Pzeta:FlatWell:final} with numerical computations, one needs to specify the potential outside the well (so one can evaluate $\calP_{\zeta,\ucl}$ appearing in the last term). For simplicity, let us consider the case of a quadratic potential interrupted by a flat quantum well,
\bea
\label{candidate1}
	V(\phi) = 
	\begin{cases}
	\frac{1}{2}m^2\phi_\uw^2 
	& \quad \mathrm{for} \quad 
	\phi_\uend \leq \phi \leq \phi_\uw 
	\\
	\frac{1}{2}m^2\phi^2 
	& \quad \mathrm{for} \quad 
	\phi \geq \phi_\uw 
	\end{cases}
	\, .
\eea
We also assume that there is no second classical phase, so $\phi_\uc=\phi_\uend$ (otherwise $\ln(k)$ is simply shifted by a constant). From \Eq{eq:meanN:class}, in the classical branch, one has $N_{\ucl,1}(\phi_*)=(\phi_*^2-\phi_\uw^2)/(4\Mp^2)$, and from \Eq{eq:Pzeta:classical:SingleField:SlowRoll}, $\calP_{\zeta,\ucl}(\phi_*)=m^2\phi_*^4/(96\pi^2\Mp^6)$. 

\begin{figure}[t]
\begin{center}
\includegraphics[width=0.48\textwidth]{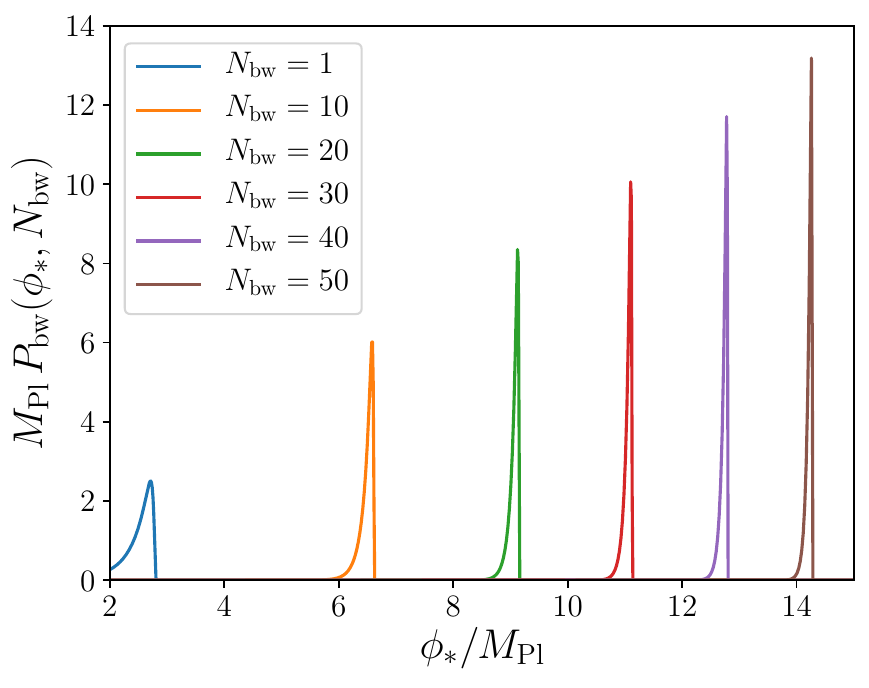}
\includegraphics[width=0.469\textwidth]{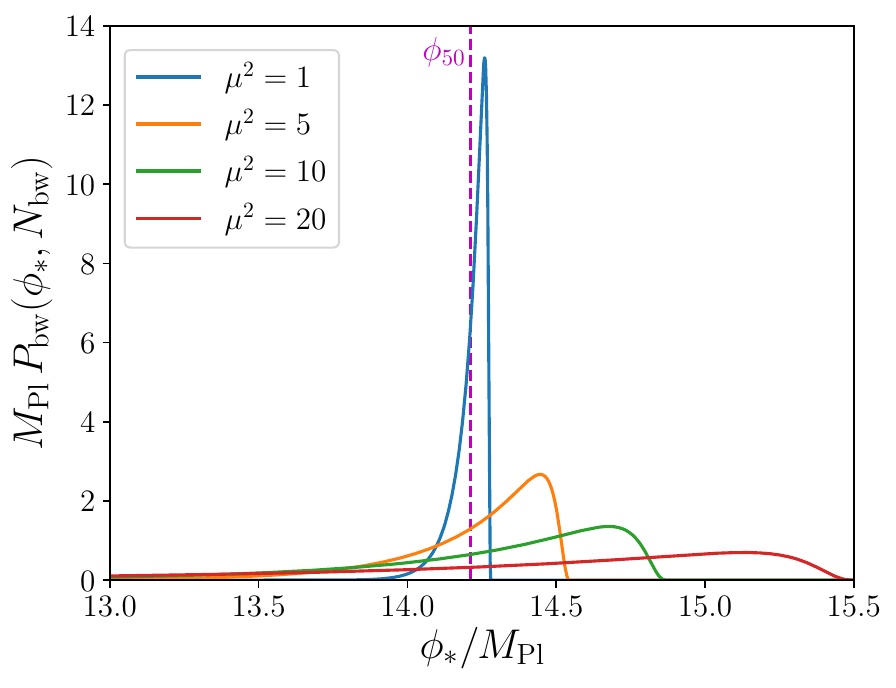}
\caption{
Backward probability outside the flat well in the quadratic potential~\eqref{candidate1}, where $m$ and $\phi_\uw$ are set according to the discussion around \Eq{param_choice_phi_w}. In the left panel, $\mu=1$ (which corresponds to $\phi_\uw=2\Mp$) and the result is shown for several values of $N_\ubw$ (the backward probability inside the well for that same value of $\mu$ is shown in \Fig{fig:Pbw_flat_reflect_mu2}). When $N_\mathrm{bw} = 1$, the probability to lie in the quantum well is $p_\uwell\simeq 0.11$, see \Eq{eq:pwell}, and this probability is negligible for the other values of $N_\mathrm{bw} = 1$. In the right panel, we set $N_\mathrm{bw} = 50$ and let $\mu$ vary. The magenta line shows the value of $\phi_*$ from which the mean number of \efolds~is $50$, \ie such that $\langle \mathcal{N} \rangle(\phi_*)=50.$}  
\label{fig:Pbw_quad}
\end{center}
\end{figure}

The model contains three parameters, namely $m$, $\mu$ and $\phi_\uw$. Let us fix two of them such that we only have one parameter to vary. In order to allow for a direct comparison with the standard case, where no quantum well is considered, we take for $m$ the standard value 
\bea
    m\simeq 6.98 \times 10^{-6}\Mp\, \frac{2\times 50 + 1}{2N_\ubw(k_\mathrm{p}) + 1} \,,
\eea
which leads to the correct normalisation of the amplitude of the power spectrum at CMB scales if the pivot scale of the CMB, $k_\mathrm{p}=0.05\,\Mpc^{-1}$, exits the Hubble radius $N_\ubw(k_\mathrm{p})$ \efolds~before the end of inflation~\cite{Martin:2013tda}. 
Furthermore, since the mean number of \efolds~elapsed from $\phi_*>\phi_\uw$ to the end of inflation is given by
\bea
	\vev{\mathcal N(\phi_*)} = N_\mathrm{cl,1}(\phi_*) + \vev{\mathcal N(\phi_\mathrm{w})}
	=\frac{\phi_*^2 - \phi_\mathrm{w}^2}{4M_\uPl^2} + \frac{\mu^2}{2}
	\, ,
\eea
let us choose $\phi_\uw$ such that this also matches the standard formula $N(\phi_*)=(\phi_*^2-\phi_\uend^2)/(4\Mp^2)$ where $\phi_\uend=\sqrt{2}\Mp$~\cite{Martin:2013tda}, which leads to 
\bea
\label{param_choice_phi_w}
	\phi_\mathrm{w} = \sqrt{2\prn{\mu^2 + 1}}\Mp
	\,.
\eea
This fixes $m$ and $\phi_\uw$, and we can keep $\mu$ as the only free parameter. When $\mu=0$, $\phi_\uw=\phi_\uc=\phi_\uend$ and one recovers standard quadratic inflation, while $\mu>0$ should yield finite stochastic corrections to the power spectrum at all scales. Note that quadratic inflation is known to yield a tensor-to-scalar ratio that is in strong tension with the latest observations~\cite{Akrami:2018odb}, but here it is used only as a toy example to illustrate the impact of quantum diffusion at small scales on scalar fluctuations (more precisely on the amplitude and spectral tilt of the scalar power spectrum) at large scales.

In \Fig{fig:Pbw_quad}, we first display the backward probability outside the well, in the first classical region. In the left panel, we set $\mu=1$ [so $\phi_\uw=2\Mp$ because of \Eq{param_choice_phi_w}], and we recall that the backward probability inside the well for that value of $\mu$ was shown in \Fig{fig:Pbw_flat_reflect_mu2}. When $N_\ubw$ increases, the distribution becomes more peaked, and centred around larger values of $\phi_*$. 
This is because the velocity of the inflaton measured in terms of {\efolds}, $\vert \dd\phi / \dd N \vert \propto 1/\phi$, is larger for smaller $\phi$.
In the right panel, we set $N_\ubw=50$ and let $\mu$ vary. Because of the choice made in \Eq{param_choice_phi_w}, the value of $\phi_*$ from which the mean number \efolds~equals $50$ is the same for all curves, and one can see that as $\mu$ decreases, the backward probability becomes more and more peaked around this value. It is also interesting to note that the backward distribution is rather skewed, and has a heavier tail at small values of $\phi_*$ (\ie towards the location of the flat well) than at large values of $\phi_*$.

\begin{figure}[t]
\begin{center}
\includegraphics[width=0.49\textwidth]{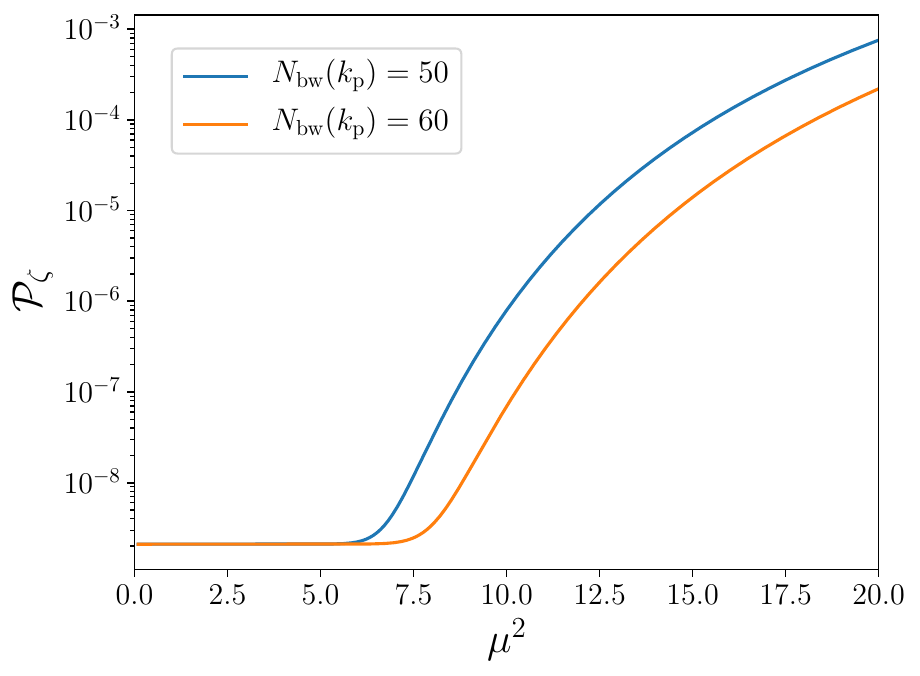}
\includegraphics[width=0.49\textwidth]{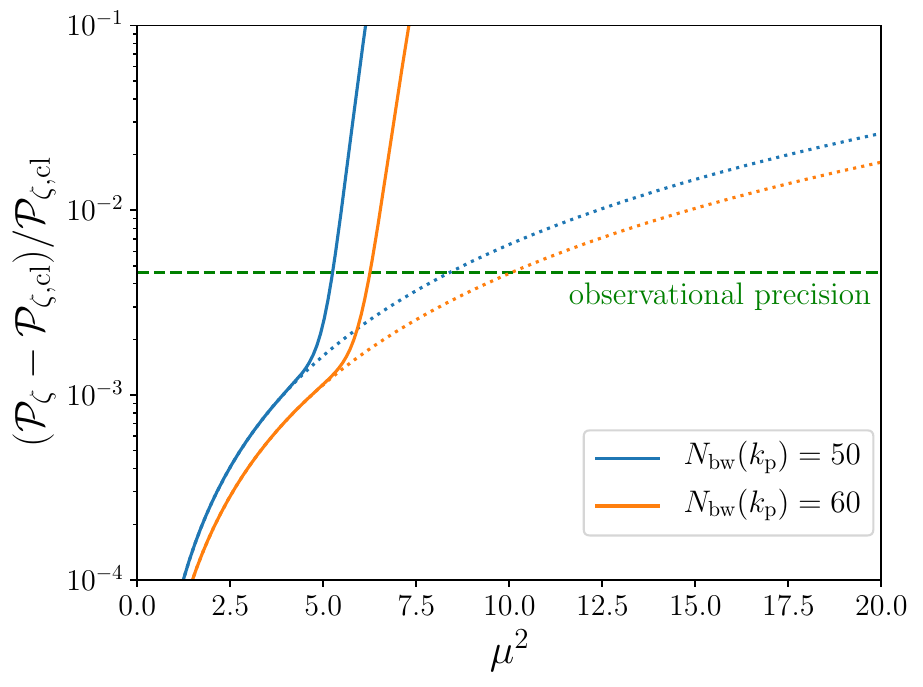}
\caption{Power spectrum of curvature perturbations at the CMB pivot scale $k_\mathrm{p}$, in a quadratic potential with a flat quantum well near the end of inflation, see \Eq{candidate1}, where $m$ and $\phi_\uw$ are set according to the discussion around \Eq{param_choice_phi_w}. The result is shown as a function of $\mu$, for $N_\ubw(k_\mathrm{p})=50$ and $N_\ubw(k_\mathrm{p})=60$. In the left panel, the absolute value of the power spectrum is displayed, while the right panel shows the relative correction to the value $\calP_{\zeta,\ucl}=2.1\times 10^{-9}$, to which the power spectrum is normalised in the limit $\mu\to 0$.
The solid lines represent the full expression while the dotted lines stand for the classical power spectrum averaged with the backward probability, \ie they correspond to the last line in \Eq{eq:Pzeta:FlatWell:final}. The green dashed line indicates the $68\%$ observational precision~\cite{Aghanim:2018eyx} on the power spectrum amplitude.}
\label{fig:diff_pZeta_ave}
\end{center}
\end{figure}
The amplitude of the power spectrum at the CMB pivot scale is then displayed as a function of $\mu$ in \Fig{fig:diff_pZeta_ave}, for $N_\ubw(k_\mathrm{p})=50$ and $N_\ubw(k_\mathrm{p})=60$. In the left panel, the absolute value of the power spectrum is shown. One can check that when $\mu\to 0$, the power spectrum approaches the observed value $\calP_{\zeta,\ucl}=2.1\times 10^{-9}$, which is guaranteed by our choice for $m$. When $\mu^2\gtrsim 6$, a substantial enhancement of the power spectrum at CMB scales is observed compared to the standard case. The correction is larger for smaller values of $N_\ubw(k_\mathrm{p})$ (the actual value of $N_\ubw(k_\mathrm{p})$ depends on the reheating dynamics~\cite{Martin:2006rs, Martin:2014nya}), given that, when $N_\ubw(k_\mathrm{p})$ decreases, the pivot scale crosses out the Hubble radius closer to the quantum well. 

In order to compare the amplitude of the stochastic corrections with observational precision, in the right panel, we show the relative difference between the full power spectrum and its classical counterpart, and we superimpose the $68\%$ observational precision from the Planck satellite measurement~\cite{Aghanim:2018eyx}. This confirms that, for $\mu^2\gtrsim 6$, the effect is within the reach of current experiments. Interestingly, $\mu^2\sim 6$ also corresponds to the point where the classical power spectrum averaged with the backward probability, \ie the term in the last line of \Eq{eq:Pzeta:FlatWell:final} [see also \Eq{eq:Power:Spectrum:Pbw:large:Phi0:LowDiffusion}], stops providing an accurate approximation to the full result. This means that, when $\mu^2\gtrsim 6$, the quantum well does not only ``blur'' the relationship between $k$ and $\phi_*$ at large scales, it also directly contributes to the power spectrum amplitude at the pivot scale.

Let us also note that the averaged classical power spectrum, \ie the last term in \Eq{eq:Pzeta:FlatWell:final}, can be approximated as follows. In the regime where it provides a good description of the full result, $p_\uwell$ has to be small, hence, from \Eq{eq:pwell}, the lower bound in the integral of the averaged power spectrum is small too. We thus assume that it can be taken to $0$, which amounts to neglecting corrections suppressed by $\ee^{- \pi^2 N_\ubw/\mu^2}$, which are of the same order as the first terms of \Eq{eq:Pzeta:FlatWell:final} that we also neglect (the fact that the function $\vartheta_2'(\pi/2,q)$ approaches $0$ when $q\to 0$ makes the approximation even better). Making use of the expressions given above for $\calP_{\zeta,\ucl}(\phi_*)$ and for $N_{\ucl,1}(\phi_*)$, the classical power spectrum as a function of $q$ is given by 
$\calP_{\zeta,\ucl}=m^2/(6\pi^2\Mp^2)[N_\ubw(k)+(\mu^2+1)/2+\mu^2 \ln(q)/\pi^2]^2$. The integral over $q$ can thus be performed exactly, by expanding the elliptic function according to its definition~\eqref{eq:elliptic:def} and by integrating each term individually, before resumming them. One obtains
\bea
\label{eq:Pzeta:averaged:quadratic}
\calP_\zeta(k)\simeq \frac{m^2}{6\pi^2\Mp^2}\left\lbrace \left[N_\ubw(k)+\frac{1}{2}\right]^2 +\frac{\mu^4}{6}\right\rbrace\, .
\eea
This provides an excellent approximation to the dotted lines in \Fig{fig:diff_pZeta_ave} (where we do not display \Eq{eq:Pzeta:averaged:quadratic} since the difference would not be seen by eye). This formula confirms that, when $\mu\to 0$, one recovers the standard result $\calP_\zeta(k)\simeq \calP_{\zeta,\ucl}(k)$. The standard result is also recovered at large values of $N_\ubw(k)$, \ie at large scales.

\begin{figure}[t]
\begin{center}
\includegraphics[width=0.49\textwidth]{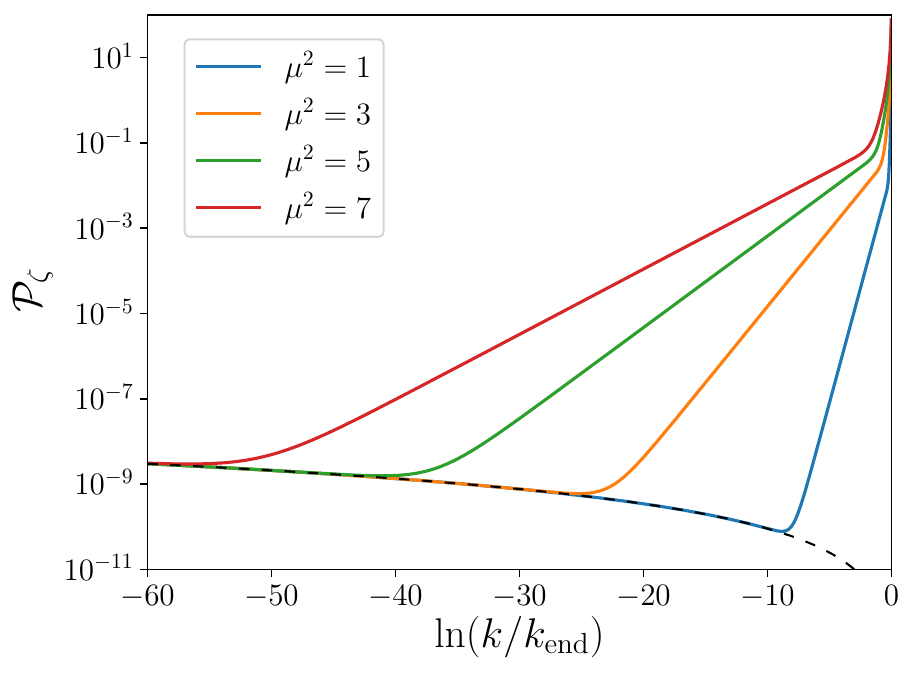}
\includegraphics[width=0.49\textwidth]{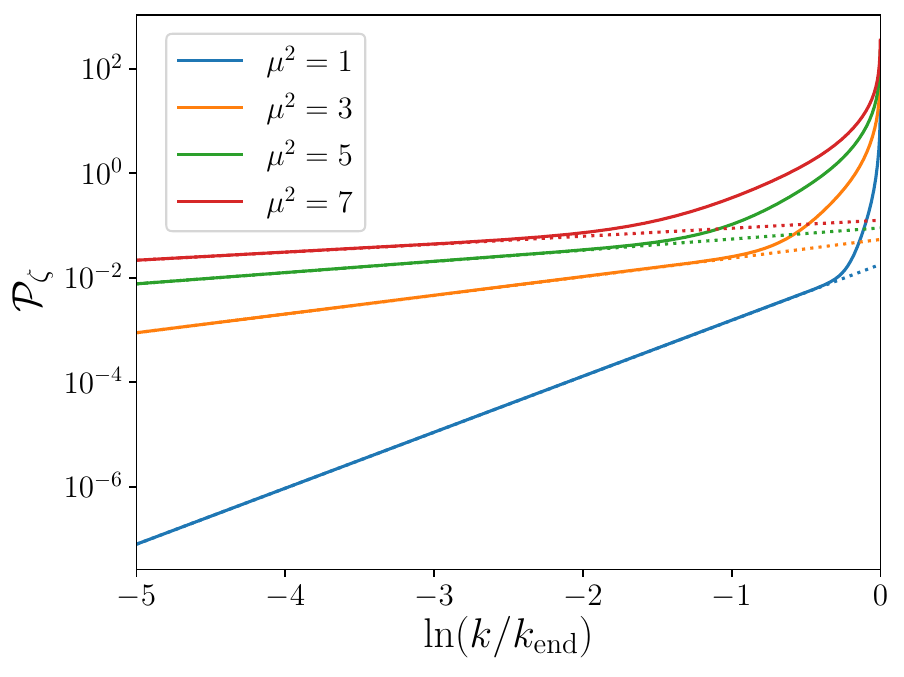}
\caption{Power spectrum of curvature perturbations in the same situation as in \Fig{fig:diff_pZeta_ave}, as a function of the wavenumber $k$, for $N_\ubw(k_\mathrm{p})=50$. In the left panel, the black dashed line stands for the classical result, see \Eq{eq:Pzeta:classical:SingleField:SlowRoll}. The right panel zooms in on the region close to $k_\uend$. The dotted lines stand for the sum of \Eqs{eq:Pzeta:averaged:quadratic} and~\eqref{eq:Pzeta:Correction}, which provides a good approximation to the full result when $N_\ubw(k)\gg \mu^2/\pi^2$.}  
\label{fig:pZeta_full}
\end{center}
\end{figure}

A shift in the overall amplitude of the power spectrum can however be easily absorbed by a change in the normalisation of $m$, hence it is not clear how the presence of those corrections would be detected experimentally. Nonetheless, we also have at our disposal high-accuracy measurements of the scale dependence of the power spectrum, in particular via its spectral index
\bea
\label{eq:SpectralIndex:def}
\nS = 1+\frac{\dd\ln\calP_\zeta}{\dd\ln k} \simeq 1-\frac{\dd\ln\calP_\zeta}{\dd N_\ubw(k)}
\eea
where in the second expression, we have made use of \Eq{eq:Nbw(k):deSitter}, which is valid at leading order in slow roll only. This is why, in \Fig{fig:pZeta_full}, we show the power spectrum as a function of $k$ for several values of $\mu^2$. The right panel zooms in on the region close to $k_\uend$. One can see that power spectrum is roughly divided into three regions. At small values of $k$, the probability to find the field inside the well $N_\ubw(k)$ \efolds~before the end of inflation is low, so the power spectrum is well approximated by the average of the classical power spectrum with the backward probability. In this regime,  \Eq{eq:Pzeta:averaged:quadratic} provides a good approximation to the full result, and from \Eq{eq:SpectralIndex:def}, the spectral index receives a perturbatively small correction,
\bea
\nS-1\simeq - \frac{2}{N_\ubw+\frac{1}{2}} \left[1-\frac{\mu^4}{\mu^4+6\left(N_\ubw+\frac{1}{2}\right)^2}\right] ,
\eea
where the first term corresponds to the standard result. At intermediate scales, this formula breaks down (as we have seen in \Fig{fig:diff_pZeta_ave}), but one can still approximate the full result by keeping the leading-order term when expanding \Eq{eq:Pzeta:FlatWell:final} in powers of $\ee^{-\pi^2 N_\ubw(k)/\mu^2}$, which leads to \Eq{eq:Pzeta:Correction}. The sum of the contributions~\eqref{eq:Pzeta:averaged:quadratic} and~\eqref{eq:Pzeta:Correction} is displayed with the dotted lines in the right panel of \Fig{fig:pZeta_full}, where one can check that it indeed provides an excellent approximation to the full result when $N_\ubw(k)\gg \mu^2/\pi^2$. In this intermediate regime, the spectral index is given by
\bea
\nS-1\simeq \frac{\pi^2}{4\mu^2}\, .
\eea
In particular, one can see that the power spectrum is blue, \ie its amplitude increases with $k$, which is ruled out by current measurements. This means that the scales observed in the CMB cannot be in the intermediate region, and this places strong constraints on $\mu$. In the present toy model, this gives rise to $\mu^2\lesssim 6$ as explained above, but this time, the effect cannot be simply re-absorbed by a change in the normalisation of the parameters of the model, since the problem comes from the colour of the spectrum. Finally, when $N_\ubw(k)\lesssim \mu^2/\pi^2$, even this approximation breaks down, and one has to make use of the full result~\eqref{eq:Pzeta:FlatWell:final}. 

Let us stress that at both intermediate and small scales, the power spectrum is found to be very blue (and even bluer at small scales than at intermediate scales). The fact that quantum diffusion breaks the quasi scale invariance of near de Sitter expansion is related to presence of the end-of-inflation surface, which acts as an absorbing wall in field space and strongly breaks field-translation invariance. This is a non trivial result, and indicates that a large enhancement of the power spectrum can arise at small scales without violating slow roll, which is otherwise often presented as a necessary condition to produce primordial black holes. 

One may be concerned that as $k$ approaches $k_\uend$, the power spectrum seems to diverge in \Fig{fig:pZeta_full}. This is however an artefact of the simple toy model we have considered in this discussion, where the slope of the potential is discontinuous as one approaches $\phi_\uc=\phi_\uend$, where inflation is ended abruptly. One may expect that, if the transition at the end of inflation was rather described by a smooth potential, the power spectrum would approach a finite value, the details of which depend on how inflation ends. Moreover, as pointed out in \Sec{sec:StatisticalTree}, the present calculation of the power spectrum in stochastic-$\delta N$ inflation neglects the presence of correlations between nearby patches. While this is a valid approximation at large scales, it becomes inaccurate for scales immediately above the Hubble radius, which indicates that the result may not be trusted for wavenumbers $k$ near $k_\uend$.

Before concluding this section, let us stress that the intermediate regime, where the averaged classical power spectrum fails to reproduce the full result and \Eq{eq:Pzeta:Correction} provides a reliable approximation, lies in the domain of parameter space where $N_\ubw(k)\gg \mu^2/\pi^2$, \ie is such that the probability to find the field in the well $N_\ubw$ \efolds~before the end of inflation is exponentially suppressed, see \Eq{eq:pwell}. If $p_\uwell(N_\ubw)$ is small, one may be surprised that the averaged classical formula breaks down, but the reason is that the power spectrum inside the quantum well can be much larger than the one at large-field values, as can be seen when comparing \Eqs{eq:Pzeta:averaged:quadratic} and~\eqref{eq:Pzeta:Correction}. This means that, although there are very few Hubble patches on the end-of-inflation surface for which $\phi_*$ lies inside the quantum well, those patches feature a very large power spectrum at the scale of interest, so large that it compensates for their sparse statistics. In such a case, one might expect a signal to arise in the statistics of very hot or very cold spots, that we may not be able to describe with the power spectrum only. This also allows us to stress that, even if $p_\uwell(N_\ubw)$ is small, the corrections induced by quantum diffusion at small scales on the large-scale power spectrum can be more involved than a simple re-averaging of the power spectrum, which shows the usefulness of the formalism we have developed in this work.

%%%%%%%%%%%%%%%%%%%%%%%%%%%%%%%%%%%%%%%%%%%%%%%%%%%%%
%%%%%%%%%%%%%%%%%%%%%%%%%%%%%%%%%%%%%%%%%%%%%%%%%%%%%
\section{Conclusion}
\label{sec:conclusion}
%%%%%%%%%%%%%%%%%%%%%%%%%%%%%%%%%%%%%%%%%%%%%%%%%%%%%
%%%%%%%%%%%%%%%%%%%%%%%%%%%%%%%%%%%%%%%%%%%%%%%%%%%%%
Let us summarise our main results and draw a few conclusions. In this work, we have derived the formalism required to compute the power spectrum of curvature perturbations in stochastic inflation. This relies on deriving the distribution of first passage times from a given position in field space, which had already been studied in previous works~\cite{Vennin:2015hra, Pattison:2017mbe,Ezquiaga:2019ftu}, but here, we have also accounted for the fact that the value of the fields driving inflation at the time when a given scale exits the Hubble radius is different for each realisation of the stochastic process. 

The most generic formulas are given by \Eqs{eq:Power:Spectrum:Exact} and~\eqref{eq:Power:Spectrum:Exact:FPT}, which can be used in numerical lattice simulations. In the quasi de-Sitter limit, a given scale emerges from the Hubble radius at a fixed number of \efolds~before the end of inflation, and this reduces to \Eq{eq:Power:Spectrum:Pbw}. This formula features the backward probability distribution, \ie the distribution of the fields value at a given number of \efolds~before the end of inflation. This is because we have computed the power spectrum in physical scales, as seen by a local observer. We have explained how to compute the backward probability in \Sec{sec:Pbw} in terms of the solutions of the Fokker-Planck and first-passage-time problems, see \Eq{Pbw_revised}.

We have then studied in more detail the possible corrections arising at CMB scales from quantum diffusion occurring at small scales (\ie at scales that cross out the Hubble radius close to the end of inflation). We have found that, in some regime, the full power spectrum is well approximated by the standard result if averaged with the backward distribution. We have however stressed that this is not always the case, and that this approximation can break down even in regimes where the probability to find the fields inside the stochastically dominated region $50$ \efolds~before the end of inflation is small. This shows that, in general, quantum diffusion at small scales does not only blur the classical relationship between wavenumbers and field values, and that it can go beyond a simple re-averaging of the standard formulas (which otherwise yield small corrections for quasi scale-invariant power spectra, \ie corrections that are proportional to the spectral running). 

Finally, for illustration, we have considered the case of a single-field quadratic potential containing a flat potential well near the end of inflation. The amplitude of the stochastic corrections in that case is controlled by the parameter denoted $\mu^2$, see \Eq{eq:Langevin:FlatPot}, which corresponds to the squared width of the well divided by its potential height, in Planckian units. We have found that when $\mu^2\gtrsim 6$, the corrections coming from the quantum well are so large that they make the power spectrum blue (\ie the spectral index is larger than one), which is excluded by CMB measurements. Let us stress that for $\mu^2=6$, the scale that crosses out the Hubble radius $50$ \efolds~before the end of inflation has probability $p_\uwell\simeq 1.5\times 10^{-9}$ to do so when the field is inside the quantum well. The CMB therefore scans regions of the inflationary potential that are very far from the quantum well, but yet, the contribution from the well to the observed power spectrum is substantial. 

This shows that, for models featuring large quantum diffusion at small scales, it is important to employ the formalism developed in this work to compute the power spectrum at large scales, even if CMB scales seem a priori (and, we argue, wrongly) immune to physics at much smaller scales. This is typically the case in models leading to primordial black holes. This also indicates that CMB measurements have the ability to constrain the shape of the inflationary potential in the entire range from the point where CMB scales are generated down to the end of inflation. 

We also found that, at small scales where quantum diffusion dominates, the power spectrum becomes very blue tilted. Contrary to the standard lore, slow-roll violations are therefore not necessary to enhance perturbations at small scales and produce primordial black holes.

Let us also mention that the constraint $\mu^2\lesssim 6$ may not seem very competitive since it was shown in \Refa{Pattison:2017mbe} that $\mu^2\gtrsim 1$ leads to an overproduction of primordial black holes anyway. However, primordial black holes arising from scales that emerge close to the end of inflation are often very light, and can Hawking evaporate before big-bang nucleosynthesis, in which case they cannot be constrained without making further assumptions (with the exception of via the stochastic gravitational wave background that is emitted in the transient phase during which they dominate the universe content~\cite{Papanikolaou:2020qtd}). More precisely, if inflation proceeds at an energy density of $10^{16} \GeV$ (which is the largest value that is compatible with current upper bounds on the tensor-to-scalar ratio in single-field slow-roll models), the Hubble mass at the end of inflation is of the order of $10\,\mathrm{g}$, and the mass that would evaporate at big-bang nucleosynthesis, $10^9\,\mathrm{g}$, crosses out the Hubble radius around $10$ \efolds~before the end of inflation if black holes form during a radiation era, and around $6$ \efolds~if they form in a matter era. From \Eq{eq:meanN:well}, the mean number of \efolds~spent in the quantum well is $\mu^2/2$, so it means that the (potentially overproduced) black holes evaporate before big-bang nucleosynthesis and can therefore not be excluded as soon as $\mu^2\lesssim20$ (if black holes form in a radiation era) or $\mu^2\lesssim12$ (if black holes form in a matter era). The constraint $\mu^2\lesssim6$ is therefore competitive. Furthermore, we have found that inside the quantum well, the power spectrum is blue-tilted. In such cases, it has recently been shown~\cite{MoradinezhadDizgah:2019wjf} that the mass distribution of primordial black holes peaks at the smallest masses, which are the ones that Hawking evaporate first. In this case, the overproduction problem may be even easier to evade, and our constraint becomes even more important. Let us also mention the possibility of having multiple quantum wells, each compatible with primordial black holes constraints, but collectively leading to a large $\mu^2$. Such models have recently been proposed in the context of the swampland conjectures~\cite{Ooguri:2006in,Obied:2018sgi,Garg:2018reu,Ooguri:2018wrx}, and primordial black holes arising from mutistaged inflation have been studied in \Refa{Tada:2019amh}.

The formalism we have developed also opens up a few prospects. First, while the concrete example we have considered in \Secs{sec:QuantumWell:Sandwiched} and~\ref{sec:Flat:Well} is the one of a single-field, slow-roll toy model, the formalism we have developed in \Sec{sec:PS} applies to any inflationary setup. In particular, we have shown that the dependence on the (probability distribution of the) initial condition $\boldmathsymbol{\Phi}_0$ drops only in single-field models in the presence of a dynamical attractor such as slow roll, but remains otherwise. It would be interesting to study the role played by initial conditions in multiple-field models.

Second, most analytical expressions we have derived are valid in the quasi de-Sitter limit, since they assume that a given scale crosses out the Hubble radius at a fixed number of \efolds~before the end of inflation (hence that $H$ is almost constant). Otherwise, analytical calculations seem difficult to carry out. However, fully numerical approaches can still be designed, making use of the exact formula~\eqref{eq:Power:Spectrum:Exact} or with \Eq{eq:Power:Spectrum:Exact:FPT}. This may be required to compute slow-roll corrections at next-to-leading order. Lattice simulations could also include the finite spatial correlation between the noise realisations in two separate Hubble patches. 

Third, the power spectrum we have computed in this work is the statistical expectation value of the power spectrum inside a given observable universe, but it should be possible to extend our formalism to compute higher moments, such as the stochastic variance of the power spectrum for instance. This would lead to a calculation of cosmic variance that includes quantum diffusion effects. One may also want to resolve the full distribution function associated to the power spectrum. In the toy-model example mentioned above, indeed, a very small fraction of the Hubble patches on the end-of-inflation surface receive a direct contribution from the quantum well, but this contribution is so large that it leads to a substantial enhancement of the mean power spectrum. Instead of the mean power spectrum, a more relevant calculation may therefore be the probability that such a patch lies inside our observable universe, or the density of the hot (or cold) spots associated to such rare patches. 

More generally, note that, in this work, as a first step, we have only performed the calculation of the power spectrum of curvature perturbations $\zeta$ at a given scale $k$, leaving the investigation of the full statistics of $\zeta_k$, and possibly, of joint distributions between the curvature perturbation at several scales, for future works. The reason is that the main application we have considered is the imprint left by quantum diffusion at small scales on large-scales observables such as the CMB. Since, at large scales, non-Gaussianities have not been detected yet, the power spectrum is the only observable we have at our disposal, which explains why we focused on the power spectrum. We however think that the tools we have developed here lay the ground for such extensions to be carried out in the future.

%%%%%%%%%%%%%%%%%%%%%%%%%%%%%%%%%%
%%%%%%%%%%% Acknowledge %%%%%%%%%%%
%%%%%%%%%%%%%%%%%%%%%%%%%%%%%%%%%%
\section*{Acknowledgements}
%\small\noindent
It is a pleasure to thank Yuichiro Tada and David Wands for interesting discussions.
K.A. is supported by JSPS Research Fellowships for Young Scientists Grant No. 18J21906, and Advanced Leading Graduate Course for Photon Science.
%\normalsize

%%%%%%%%%%%%%%%%%%%%%%%%%%%%%%%%%%
%%%%%%%%%%% Appendix %%%%%%%%%%%
%%%%%%%%%%%%%%%%%%%%%%%%%%%%%%%%%%

\appendix

\section{Solutions to the Fokker-Planck and adjoint Fokker-Planck equations in a flat potential well}
\label{sec:FP:sol:Flat}
In this appendix, we derive the solution of the Fokker-Planck equation for the distribution function of the field, $P^\uwell(\phi,N)$, and of the adjoint Fokker-Planck equation for the distribution function of first passage times, $\Pfpt(\mathcal{N},\phi)$, in a flat potential well. Following the discussion presented in \Sec{sec:Flat:Well}, we label the field with the variable $x$ introduced in \Eq{eq:Langevin:FlatPot}, in terms of which the Langevin equation reads
\bea
\label{eq:Langevin:FlatPot:App}
\frac{\dd x}{\dd N} = \frac{\sqrt{2}}{\mu}\xi(N)\, .
\eea 
Inside the well, the variable $x$ varies between $0$, where an absorbing boundary is located, and $1$, where a reflective boundary is placed. The Fokker-Planck equation associated to \Eq{eq:Langevin:FlatPot:App} is given by~\cite{risken1989fpe}
\bea
\label{eq:FokkerPlanck:FlatPot:App}
\frac{\partial}{\partial N} P^\uwell(x,N\vert x_\uin,0) = \frac{1}{\mu^2}\frac{\partial^2}{\partial x^2} P^\uwell(x,N\vert x_\uin,0)\, .
\eea
This result does not depend on the discretisation scheme the Langevin equation is interpreted with, see footnote~\ref{footnote:Discretisation:Strato:vs:Ito}.
%%%%%%%%%%%%%%%%%%%%%%%%%%%%%%%%%%%%%%%%%%%%%%%%%%%%%
\subsection{Flat well with one absorbing wall}
\label{subsec:Pbw_revised_flat_abs_wall1}
%%%%%%%%%%%%%%%%%%%%%%%%%%%%%%%%%%%%%%%%%%%%%%%%%%%%%
As a warm up, let us first consider the case where only the absorbing boundary at $x=0$ is considered. This boundary condition imposes that $ P^\uwell(x,N\vert x_\uin,0)=0$ when $x=0$, and we set the initial condition at $x_\uin$ when $N=0$, \ie $ P^\uwell(x,0\vert x_\uin,0)=\delta(x-x_\uin)$. A first remark is that \Eq{eq:FokkerPlanck:FlatPot:App} describes free Brownian motion with diffusivity $2/\mu^2$, hence it accepts Gaussian solutions of the form
\bea
\label{eq:FP:sol:fundamental}
f_{\bar{x},\bar{N}}(x,N)=\frac{\mu}{2\sqrt{\pi \left(N-\bar{N}\right)}}\ee^{-\frac{\mu^2}{4}\frac{\left(x-\bar{x}\right)^2}{N-\bar{N}}}\, ,
\eea
where $\bar{x}$ and $\bar{N}$ are two integration constants. Since the Fokker-Planck operator is linear and second order, those functions form a basis on which all solutions can be expanded. In practice, the coefficients in these expansions are set in order to satisfy the boundary conditions. Let us note that in the absence of any boundary condition, the solution would be simply given by $P^\uwell(x,N\vert x_\uin,0)=f_{x_\uin,0}(x,N)$. However this solution does not satisfy the boundary condition at $x=0$. This can be fixed by subtracting the same solution, but centred at the symmetric point $-x_\uin$, which gives rise to $P^\uwell(x,N\vert x_\uin,0)=f_{x_\uin,0}(x,N)-f_{-x_\uin,0}(x,N)$, or more explicitly,
\bea
\label{eq:Pwell:OneAbsorbing}
P^\uwell(x,N\vert x_\uin,0)=\frac{\mu}{2\sqrt{\pi N}}\left[\ee^{-\frac{\mu^2}{4}\frac{\left(x-x_\uin\right)^2}{N}}-\ee^{-\frac{\mu^2}{4}\frac{\left(x+x_\uin\right)^2}{N}}\right]\theta(x)\, .
\eea
By construction, this function satisfies the Fokker-Planck equation, it vanishes at $x=0$, and it is such that $P^\uwell(x,0\vert x_\uin,0)=[\delta(x-x_\uin)-\delta(x+x_\uin)]\theta(x)=\delta(x-x_\uin)$. It has been obtained by removing the mirrored image of the solution obtained without boundary conditions (where the ``mirror'' is thought of as sitting at the location of the absorbing boundary), and for that reason this technique is often referred to as the ``method of images''.

The distribution associated to first passage times through the absorbing boundary can then be obtained as follows. On the one hand, from the solution of the Fokker-Planck equation, one can compute the survival probability $S(N)$, which corresponds to the probability that the field is still within the well at time $N$,
\bea
S(N)=\int_0^\infty  P^\uwell(x,N\vert x_\uin,0) \dd x\, .
\eea
On the other hand, the probability to have already escaped from the well at time $N$ is given by 
\bea
\int_0^N \Pfpt(\mathcal{N},x_\uin)\dd\mathcal{N}=1-S(N)\, .
\eea
By differentiating both expressions with respect to $N$, one obtains
\bea
\Pfpt(N,x_\uin) = - \frac{\partial}{\partial N} \int_0^\infty  P^\uwell(x,N\vert x_\uin,0) \dd x\, .
\eea
One can then use the fact that $P^\uwell $ satisfies the Fokker-Planck equation~\eqref{eq:FokkerPlanck:FlatPot:App}, which leads to
\bea
\label{eq:Pfpt:from:Pprime}
\Pfpt(N,x_\uin) = \frac{1}{\mu^2} 
\left.\frac{\partial}{\partial x}P^\uwell(x,N\vert x_\uin,0) \right\vert_{x=0}\, .
\eea
Making use of \Eq{eq:Pwell:OneAbsorbing}, one finally obtains a Levy distribution, 
\bea
\label{FPT_density_flat_single}
\Pfpt(N,x_\uin) =\frac{\mu x_\uin}{2\sqrt{\pi}N^{3/2}}\ee^{-\frac{\mu^2 x_\uin^2}{4 N}}\, .
\eea

\begin{figure}[t]
\begin{center}
\includegraphics[width=0.5\textwidth]{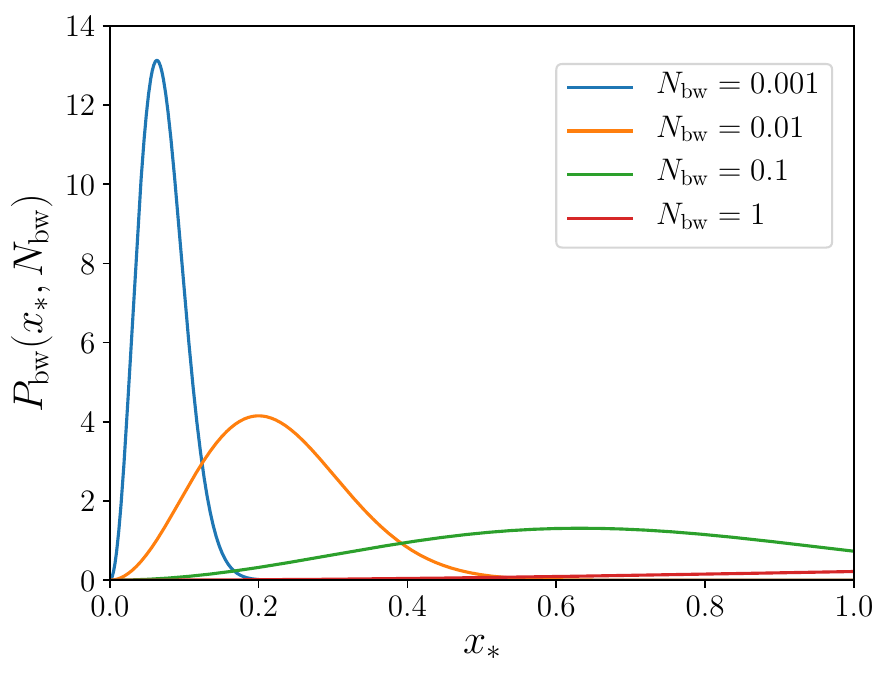}
\caption{Backward probability distribution for a flat potential with an absorbing boundary at $x= 0$ and in the $x_0\rightarrow\infty$ limit, computed with \Eq{Pbw_flat_single_revised_lim}, for $\mu=1$ and with four different values of $N_\ubw$.
}  
\label{fig:Pbw_flat_single_right}
\end{center}
\end{figure}
In passing, let us compute the backwards probability distribution in this semi-infinite well. This can be done by plugging \Eqs{eq:Pwell:OneAbsorbing} and~\eqref{FPT_density_flat_single} into \Eq{Pbw_revised}. The integral of ${\Pfpt}(N,\phi_0)$ over $N$ can be performed 
by means of the error function,
\bea
\label{FPT_density_flat_single_integrate}
	\int_{N_\mathrm{bw}}^\infty \dd N \,
	{\Pfpt}\left(N, x_0 \right)
	= \erf\prn{\frac{x_0\mu}{2\sqrt{ N_\mathrm{bw} }}} \,.
\eea
This approaches one when $x_0\to \infty$, in agreement with the discussion below \Eq{Pbw_revised}. The integral of $P^\uwell(x,N\vert x_0,0)$ over $N$ also takes a simple form, and combining the above results, one obtains
\bea
P_\ubw\left(x_*,N_\ubw\right)=\frac{\mu^3 x_* \min(x_*,x_0)}{2\sqrt{\pi}N_\ubw^{3/2}}
\frac{\ee^{-\frac{\mu^2 x^2}{4 N_\ubw}}}{\erf\left(\frac{x_0\mu}{2\sqrt{N_\ubw}}\right)}\, .
\eea
One can check explicitly that this distribution is properly normalised, and that it vanishes when $x_*=0$. In the limit where $x_0$ is sent to infinity, it reduces to 
\bea
\label{Pbw_flat_single_revised_lim}
P_\ubw\left(x,N_\ubw\right)=\frac{\mu^3 x_*^2}{2\sqrt{\pi}N_\ubw^{3/2}}
\ee^{-\frac{\mu^2 x_*^2}{4 N_\ubw}}\, .
\eea
This distribution is displayed in \Fig{fig:Pbw_flat_single_right} for $\mu=1$ 
and for a few values of $N_\ubw$. One can check that, for small values of $N_\ubw$, the distribution is peaked close to $x=0$, while it is more widely spread and centred around larger values of $x$ for larger values of $N_\ubw$. 
%%%%%%%%%%%%%%%%%%%%%%%%%%%%%%%%%%%%%%%%%%%%%%%%%%%%%
\subsection{Flat well with one absorbing wall and one reflective wall}
\label{subsec:Pbw_revised_flat_reflect}
%%%%%%%%%%%%%%%%%%%%%%%%%%%%%%%%%%%%%%%%%%%%%%%%%%%%%
Let us now study the setup of interest for \Sec{sec:Flat:Well}, where, on top of the absorbing boundary at $x=0$, one places a reflective boundary at $x=1$. Compared to the situation studied in \Sec{subsec:Pbw_revised_flat_abs_wall1}, the equations are slightly more involved, but the calculational techniques are the same, which is why it was useful to first study the simple case where only the absorbing boundary is accounted for. 

The method of images can be employed by introducing the image sources at locations
\bea
\label{eq:sources:location}
\bar{x}_n=n+\frac{1}{2}+(-1)^n\left(x_\uin-\frac{1}{2}\right)\, .
\eea
When $n=0$, one has $\bar{x}_n=x_\uin$, so one initiates this set of image sources at the initial location of the field. Then, when mirrored through the absorbing wall, one has $\bar{x}_n \to - \bar{x}_n=\bar{x}_{-n-1}$, and when mirrored through the reflective wall, one has $\bar{x}_n\to 2-\bar{x}_n=\bar{x}_{-n+1}$. Therefore, the values $\bar{x}_n$ indeed correspond to all possible image sources of $x_\uin$ after an arbitrary number of reflections against the absorbing and reflective walls. The solution to the Fokker-Planck equation should be constructed as a sum of free solutions centred on each of these images, with a sign that remains the same when going through the reflective wall and that flips when going through the absorbing wall. Starting from $x_\uin$, schematically, one has
\bea
\underbrace{x_\uin}_{+}   \overset{R}{\longrightarrow} \underbrace{\bar{x}_1}_{+}   \overset{A}{\longrightarrow} \underbrace{\bar{x}_{-2}}_{-} \overset{R}{\longrightarrow} \underbrace{\bar{x}_3}_{-}\overset{A}{\longrightarrow} \underbrace{\bar{x}_{-4}}_{+}  \overset{R}{\longrightarrow} \underbrace{\bar{x}_5}_{+}          \longrightarrow \cdots\\
\underbrace{x_\uin}_{+}   \overset{A}{\longrightarrow} \underbrace{x_{-1}}_{-}   \overset{R}{\longrightarrow} \underbrace{\bar{x}_{2}}_{-} \overset{A}{\longrightarrow} \underbrace{\bar{x}_{-3}}_{+}\overset{R}{\longrightarrow} \underbrace{\bar{x}_{4}}_{+}  \overset{A}{\longrightarrow} \underbrace{\bar{x}_{-5}}_{-}          \longrightarrow \cdots
\eea
where $A$ and $R$ mean reflection against the absorbing and reflective walls respectively, and the sign associated to each image source is displayed below it. One can see that it is given by $({-1})^{\lfloor n/2 \rfloor}$, where $\lfloor \cdot \rfloor$ denotes the floor integer part. One thus has
\bea
P^\uwell(x,N\vert x_\uin,0) =\theta(x)\theta(1-x)\sum_{n=-\infty}^\infty ({-1})^{\lfloor \frac{n}{2} \rfloor} f_{\bar{x}_n,0} \left(x,N\right)\, .
\eea
Hereafter, we will drop the Heaviside functions and remember that the above expression applies within the well only. It can be decomposed into two sums, one where $n=2m$ is even and one where $n=2m+1$ is odd. In both cases, $\lfloor n/2 \rfloor=m$, and making use of \Eq{eq:FP:sol:fundamental}, this gives rise to
\bea
\kern-2em
	P^\uwell(x,N\vert x_\uin,0)=\frac{\mu}{2\sqrt{\pi N}} 
	\com{
	\sum_{m=-\infty}^\infty (-1)^m \ee^{-\frac{\mu^2\left(x-\bar{x}_{2m}\right)^2}{4N}}
	+ \sum_{m=-\infty}^\infty (-1)^m \ee^{-\frac{\mu^2\left(x-\bar{x}_{2m+1}\right)^2}{4N}}
	} 	\,.
\eea
If one replaces the locations of the image sources by their expression~\eqref{eq:sources:location}, one obtains
\bea
P^\uwell(x,N\vert x_\uin,0)=\frac{\mu}{2\sqrt{\pi N}} \sum_{m=-\infty}^\infty (-1)^m \ee^{-\frac{\mu^2}{4N}\left(x-x_\uin-2m\right)^2}-\lbrace x_\uin \to -x_\uin \rbrace\, .
\eea
By expanding the square in the argument of the exponential function, one notices that the result can be written in terms of the fourth elliptic theta function $\vartheta_4$,
\bea
P^\uwell(x,N\vert x_\uin,0)=\frac{\mu}{2\sqrt{\pi N}}\ee^{-\frac{\mu^2}{4N}\left(x-x_\uin\right)^2}\vartheta_4\left[\frac{i\mu^2}{2N}\left(x-x_\uin\right),\ee^{-\frac{\mu^2}{N}}\right]-\lbrace x_\uin \to -x_\uin \rbrace\, ,
\label{eq:Pwell:flat:interm}
\eea
where $\vartheta_4$ is defined in \Eq{eq:elliptic:def}, see footnote~\ref{footnote:elliptic:def} in the main text. The next step is to notice that this expression can be further simplified by making use of the Jacobi identity 
\bea
\vartheta_4\left(iz,\ee^{-\pi\tau}\right) = \frac{\ee^{\frac{z^2}{\pi\tau}}}{\sqrt{\tau}}\vartheta_2\left(-\frac{z}{\tau},\ee^{-\frac{\pi}{\tau}}\right)\, ,
\eea
see Eq.~(20.7.33) of \Refa{2011ConPh..52..497T}, where $\vartheta_2$ is defined in \Eq{eq:elliptic:def}. This allows one to rewrite \Eq{eq:Pwell:flat:interm} as 
\bea
\label{eq:Pwell:flat:App}
P^\uwell(x,N\vert x_\uin,0)= \frac{1}{2}\vartheta_2\left[-\frac{\pi}{2}\left(x-x_\uin\right),\ee^{-\frac{\pi^2 N}{\mu^2}}\right]-\frac{1}{2}\vartheta_2\left[-\frac{\pi}{2}\left(x+x_\uin\right),\ee^{-\frac{\pi^2 N}{\mu^2}}\right]\, .
\eea
This expression also allows one to compute the distribution of first passage times, making use of \Eq{eq:Pfpt:from:Pprime}. The result is given in terms of the $\vartheta_2'$ function, which denotes the derivative of $\vartheta_2$ with respect to its first argument. Since $\vartheta_2$ is an even function of its first argument, $\vartheta_2'$ is an odd function, so the two terms in \Eq{eq:Pwell:flat:App} provide the same contribution and one obtains
\bea
\Pfpt^\uwell(x,N)=-\frac{\pi}{2\mu^2}\vartheta_2'\left(\frac{\pi}{2}x,\ee^{-\frac{\pi^2 N}{\mu^2}}\right)\, .
\eea
This matches Eq.~(4.11) of \Refa{Pattison:2017mbe}, although here it is derived using different techniques.

Finally, let us note that in order to evaluate the backward probability with \Eq{Pbw_revised}, one needs to integrate \Eq{eq:Pwell:flat:App} over $N$ when setting $x_\uin=1$. A first remark is that, making use of the definition of the elliptic functions, see \Eq{eq:elliptic:def}, one can easily show that $\vartheta_2(z\pm\pi/2,q)=\pm\vartheta_1(-z,q)$. As a consequence, one has
\bea
P^\uwell(x,N\vert 1,0)=\vartheta_1\left(\frac{\pi}{2}x,\ee^{-\frac{\pi^2 N}{\mu^2}}\right)\, .
\eea
After performing the change of integration variable $\ee^{-\pi^2 N/\mu^2} = q$, and denoting $z=\pi x/2$, one thus has to compute
\bea
\int_0^\infty  P^\uwell(x,N) \dd N = \frac{\mu^2}{\pi^2}\int_0^1 \frac{\dd q}{q} \vartheta_1\left(z,q\right)\, .
\eea
By expanding the $\vartheta_1$ function according to its definition given in \Eq{eq:elliptic:def}, each term can be integrated exactly, and this gives rise to
\bea
\int_0^1 \frac{\dd q}{q} \vartheta_1\left(\frac{\pi}{2}x,q\right) = 8 \sum_{n=0}^\infty (-1)^n \frac{\sin\left[\left(2n+1\right)z\right]}{(2n+1)^2}\equiv F(z)\, ,
\eea
which defines the function $F(z)$. By differentiating this function, one obtains
\bea
F'(z)&=& 8 \sum_{n=0}^\infty (-1)^n\frac{\cos\left[\left(2n+1\right)z\right]}{(2n+1)}\\
& =& 4 \left[\underbrace{\sum_{n=0}^\infty \frac{(-1)^n}{2n+1}\ee^{(2n+1)i z}}_{\mathrm{arctan}\left(\ee^{i z}\right)}+\underbrace{\sum_{n=0}^\infty \frac{(-1)^n}{2n+1}\ee^{-(2n+1)i z}}_{\mathrm{arctan}\left(\ee^{-i z}\right)}\right]\, ,
\eea
where we one recognises the Taylor expansion of the $\mathrm{arctan}$ function. The next step is to make use of the identity $\mathrm{arctan}(Z)+\mathrm{arctan}(1/Z)=\pi/2$, which is valid when the argument of $Z$ is comprised between $0$ and $\pi/2$, which is the case here since $Z=\ee^{i z}=\ee^{i\pi x/2}$ and $0\leq x\leq 1$. This gives rise to $F'(z)=2\pi$. Furthermore, when $z=0$, $\theta_1(z,q)=0$ so $F(0)=0$ and one obtains $F(z)=2\pi z$. Combining the above results, one is led to
\bea
\int_0^\infty  P^\uwell(x,N) \dd N = \mu^2 x\, ,
\eea
which we use in the main text below \Eq{eq:deltaN2:well}.
\bibliographystyle{JHEP}
\bibliography{StocPzetak}

\providecommand{\href}[2]{#2}\begingroup\raggedright\begin{thebibliography}{10}

\bibitem{Starobinsky:1982ee}
A.~A. Starobinsky, \emph{{Dynamics of Phase Transition in the New Inflationary
  Universe Scenario and Generation of Perturbations}},
  \href{http://dx.doi.org/10.1016/0370-2693(82)90541-X}{\emph{Phys. Lett. B}
  {\bf 117} (1982) 175--178}.

\bibitem{Starobinsky:1986fx}
A.~A. Starobinsky, \emph{{Stochastic De Sitter (inflationary) stage in the
  early universe}},
  \href{http://dx.doi.org/10.1007/3-540-16452-9_6}{\emph{Lect. Notes Phys.}
  {\bf 246} (1986) 107--126}.

\bibitem{Polarski:1995jg}
D.~Polarski and A.~A. Starobinsky, \emph{{Semiclassicality and decoherence of
  cosmological perturbations}},
  \href{http://dx.doi.org/10.1088/0264-9381/13/3/006}{\emph{Class. Quant.
  Grav.} {\bf 13} (1996) 377--392},
  [\href{http://arxiv.org/abs/gr-qc/9504030}{{\tt gr-qc/9504030}}].

\bibitem{Lesgourgues:1996jc}
J.~Lesgourgues, D.~Polarski and A.~A. Starobinsky, \emph{{Quantum to classical
  transition of cosmological perturbations for nonvacuum initial states}},
  \href{http://dx.doi.org/10.1016/S0550-3213(97)00224-1}{\emph{Nucl. Phys. B}
  {\bf 497} (1997) 479--510}, [\href{http://arxiv.org/abs/gr-qc/9611019}{{\tt
  gr-qc/9611019}}].

\bibitem{Kiefer:2008ku}
C.~Kiefer and D.~Polarski, \emph{{Why do cosmological perturbations look
  classical to us?}}, \href{http://dx.doi.org/10.1166/asl.2009.1023}{\emph{Adv.
  Sci. Lett.} {\bf 2} (2009) 164--173},
  [\href{http://arxiv.org/abs/0810.0087}{{\tt 0810.0087}}].

\bibitem{Martin:2015qta}
J.~Martin and V.~Vennin, \emph{{Quantum Discord of Cosmic Inflation: Can we
  Show that CMB Anisotropies are of Quantum-Mechanical Origin?}},
  \href{http://dx.doi.org/10.1103/PhysRevD.93.023505}{\emph{Phys. Rev. D} {\bf
  93} (2016) 023505}, [\href{http://arxiv.org/abs/1510.04038}{{\tt
  1510.04038}}].

\bibitem{Campo:2005sv}
D.~Campo and R.~Parentani, \emph{{Inflationary spectra and violations of Bell
  inequalities}},
  \href{http://dx.doi.org/10.1103/PhysRevD.74.025001}{\emph{Phys. Rev. D} {\bf
  74} (2006) 025001}, [\href{http://arxiv.org/abs/astro-ph/0505376}{{\tt
  astro-ph/0505376}}].

\bibitem{Maldacena:2015bha}
J.~Maldacena, \emph{{A model with cosmological Bell inequalities}},
  \href{http://dx.doi.org/10.1002/prop.201500097}{\emph{Fortsch. Phys.} {\bf
  64} (2016) 10--23}, [\href{http://arxiv.org/abs/1508.01082}{{\tt
  1508.01082}}].

\bibitem{Martin:2017zxs}
J.~Martin and V.~Vennin, \emph{{Obstructions to Bell CMB Experiments}},
  \href{http://dx.doi.org/10.1103/PhysRevD.96.063501}{\emph{Phys. Rev. D} {\bf
  96} (2017) 063501}, [\href{http://arxiv.org/abs/1706.05001}{{\tt
  1706.05001}}].

\bibitem{Ando:2020kdz}
K.~Ando and V.~Vennin, \emph{{Bipartite temporal Bell inequalities for two-mode
  squeezed states}},
  \href{http://dx.doi.org/10.1103/PhysRevA.102.052213}{\emph{Phys. Rev. A} {\bf
  102} (2020) 052213}, [\href{http://arxiv.org/abs/2007.00458}{{\tt
  2007.00458}}].

\bibitem{Green:2020whw}
D.~Green and R.~A. Porto, \emph{{Signals of a Quantum Universe}},
  \href{http://dx.doi.org/10.1103/PhysRevLett.124.251302}{\emph{Phys. Rev.
  Lett.} {\bf 124} (2020) 251302}, [\href{http://arxiv.org/abs/2001.09149}{{\tt
  2001.09149}}].

\bibitem{Starobinsky:1994bd}
A.~A. Starobinsky and J.~Yokoyama, \emph{{Equilibrium state of a
  selfinteracting scalar field in the de Sitter background}},
  \href{http://dx.doi.org/10.1103/PhysRevD.50.6357}{\emph{Phys. Rev. D} {\bf
  50} (1994) 6357--6368}, [\href{http://arxiv.org/abs/astro-ph/9407016}{{\tt
  astro-ph/9407016}}].

\bibitem{Tsamis:2005hd}
N.~Tsamis and R.~Woodard, \emph{{Stochastic quantum gravitational inflation}},
  \href{http://dx.doi.org/10.1016/j.nuclphysb.2005.06.031}{\emph{Nucl. Phys. B}
  {\bf 724} (2005) 295--328}, [\href{http://arxiv.org/abs/gr-qc/0505115}{{\tt
  gr-qc/0505115}}].

\bibitem{Finelli:2008zg}
F.~Finelli, G.~Marozzi, A.~Starobinsky, G.~Vacca and G.~Venturi,
  \emph{{Generation of fluctuations during inflation: Comparison of stochastic
  and field-theoretic approaches}},
  \href{http://dx.doi.org/10.1103/PhysRevD.79.044007}{\emph{Phys. Rev. D} {\bf
  79} (2009) 044007}, [\href{http://arxiv.org/abs/0808.1786}{{\tt 0808.1786}}].

\bibitem{Garbrecht:2013coa}
B.~Garbrecht, G.~Rigopoulos and Y.~Zhu, \emph{{Infrared correlations in de
  Sitter space: Field theoretic versus stochastic approach}},
  \href{http://dx.doi.org/10.1103/PhysRevD.89.063506}{\emph{Phys. Rev. D} {\bf
  89} (2014) 063506}, [\href{http://arxiv.org/abs/1310.0367}{{\tt 1310.0367}}].

\bibitem{Vennin:2015hra}
V.~Vennin and A.~A. Starobinsky, \emph{{Correlation Functions in Stochastic
  Inflation}},
  \href{http://dx.doi.org/10.1140/epjc/s10052-015-3643-y}{\emph{Eur. Phys. J.
  C} {\bf 75} (2015) 413}, [\href{http://arxiv.org/abs/1506.04732}{{\tt
  1506.04732}}].

\bibitem{Onemli:2015pma}
V.~Onemli, \emph{{Vacuum Fluctuations of a Scalar Field during Inflation:
  Quantum versus Stochastic Analysis}},
  \href{http://dx.doi.org/10.1103/PhysRevD.91.103537}{\emph{Phys. Rev. D} {\bf
  91} (2015) 103537}, [\href{http://arxiv.org/abs/1501.05852}{{\tt
  1501.05852}}].

\bibitem{Burgess:2015ajz}
C.~Burgess, R.~Holman and G.~Tasinato, \emph{{Open EFTs, IR effects \&
  late-time resummations: systematic corrections in stochastic inflation}},
  \href{http://dx.doi.org/10.1007/JHEP01(2016)153}{\emph{JHEP} {\bf 01} (2016)
  153}, [\href{http://arxiv.org/abs/1512.00169}{{\tt 1512.00169}}].

\bibitem{Vennin:2016wnk}
V.~Vennin, H.~Assadullahi, H.~Firouzjahi, M.~Noorbala and D.~Wands,
  \emph{{Critical Number of Fields in Stochastic Inflation}},
  \href{http://dx.doi.org/10.1103/PhysRevLett.118.031301}{\emph{Phys. Rev.
  Lett.} {\bf 118} (2017) 031301}, [\href{http://arxiv.org/abs/1604.06017}{{\tt
  1604.06017}}].

\bibitem{Hardwick:2017fjo}
R.~J. Hardwick, V.~Vennin, C.~T. Byrnes, J.~Torrado and D.~Wands, \emph{{The
  stochastic spectator}},
  \href{http://dx.doi.org/10.1088/1475-7516/2017/10/018}{\emph{JCAP} {\bf 10}
  (2017) 018}, [\href{http://arxiv.org/abs/1701.06473}{{\tt 1701.06473}}].

\bibitem{Tokuda:2017fdh}
J.~Tokuda and T.~Tanaka, \emph{{Statistical nature of infrared dynamics on de
  Sitter background}},
  \href{http://dx.doi.org/10.1088/1475-7516/2018/02/014}{\emph{JCAP} {\bf 02}
  (2018) 014}, [\href{http://arxiv.org/abs/1708.01734}{{\tt 1708.01734}}].

\bibitem{Kitamoto:2018dek}
H.~Kitamoto, \emph{{Infrared resummation for derivative interactions in de
  Sitter space}},
  \href{http://dx.doi.org/10.1103/PhysRevD.100.025020}{\emph{Phys. Rev. D} {\bf
  100} (2019) 025020}, [\href{http://arxiv.org/abs/1811.01830}{{\tt
  1811.01830}}].

\bibitem{Markkanen:2019kpv}
T.~Markkanen, A.~Rajantie, S.~Stopyra and T.~Tenkanen, \emph{{Scalar
  correlation functions in de Sitter space from the stochastic spectral
  expansion}},
  \href{http://dx.doi.org/10.1088/1475-7516/2019/08/001}{\emph{JCAP} {\bf 08}
  (2019) 001}, [\href{http://arxiv.org/abs/1904.11917}{{\tt 1904.11917}}].

\bibitem{Lifshitz:1960}
E.~M. Lifshitz and I.~M. Khalatnikov, \emph{{About singularities of
  cosmological solutions of the gravitational equations. I}}, {\emph{ZhETF}
  {\bf 39} (1960) 149}.

\bibitem{Starobinsky:1982mr}
A.~A. Starobinsky, \emph{{Isotropization of arbitrary cosmological expansion
  given an effective cosmological constant}}, {\emph{JETP Lett.} {\bf 37}
  (1983) 66--69}.

\bibitem{Comer:1994np}
G.~Comer, N.~Deruelle, D.~Langlois and J.~Parry, \emph{{Growth or decay of
  cosmological inhomogeneities as a function of their equation of state}},
  \href{http://dx.doi.org/10.1103/PhysRevD.49.2759}{\emph{Phys. Rev. D} {\bf
  49} (1994) 2759--2768}.

\bibitem{Wands:2000dp}
D.~Wands, K.~A. Malik, D.~H. Lyth and A.~R. Liddle, \emph{{A New approach to
  the evolution of cosmological perturbations on large scales}},
  \href{http://dx.doi.org/10.1103/PhysRevD.62.043527}{\emph{Phys. Rev. D} {\bf
  62} (2000) 043527}, [\href{http://arxiv.org/abs/astro-ph/0003278}{{\tt
  astro-ph/0003278}}].

\bibitem{Khalatnikov:2002kn}
I.~Khalatnikov and A.~Kamenshchik, \emph{{Comment about quasiisotropic solution
  of Einstein equations near cosmological singularity}},
  \href{http://dx.doi.org/10.1088/0264-9381/19/14/322}{\emph{Class. Quant.
  Grav.} {\bf 19} (2002) 3845--3850},
  [\href{http://arxiv.org/abs/gr-qc/0204045}{{\tt gr-qc/0204045}}].

\bibitem{Lyth:2003im}
D.~H. Lyth and D.~Wands, \emph{{Conserved cosmological perturbations}},
  \href{http://dx.doi.org/10.1103/PhysRevD.68.103515}{\emph{Phys. Rev. D} {\bf
  68} (2003) 103515}, [\href{http://arxiv.org/abs/astro-ph/0306498}{{\tt
  astro-ph/0306498}}].

\bibitem{Lyth:2004gb}
D.~H. Lyth, K.~A. Malik and M.~Sasaki, \emph{{A General proof of the
  conservation of the curvature perturbation}},
  \href{http://dx.doi.org/10.1088/1475-7516/2005/05/004}{\emph{JCAP} {\bf 05}
  (2005) 004}, [\href{http://arxiv.org/abs/astro-ph/0411220}{{\tt
  astro-ph/0411220}}].

\bibitem{Starobinsky:1986fxa}
A.~A. Starobinsky, \emph{{Multicomponent de Sitter (Inflationary) Stages and
  the Generation of Perturbations}}, {\emph{JETP Lett.} {\bf 42} (1985)
  152--155}.

\bibitem{Sasaki:1995aw}
M.~Sasaki and E.~D. Stewart, \emph{{A General analytic formula for the spectral
  index of the density perturbations produced during inflation}},
  \href{http://dx.doi.org/10.1143/PTP.95.71}{\emph{Prog. Theor. Phys.} {\bf 95}
  (1996) 71--78}, [\href{http://arxiv.org/abs/astro-ph/9507001}{{\tt
  astro-ph/9507001}}].

\bibitem{Enqvist:2008kt}
K.~Enqvist, S.~Nurmi, D.~Podolsky and G.~Rigopoulos, \emph{{On the divergences
  of inflationary superhorizon perturbations}},
  \href{http://dx.doi.org/10.1088/1475-7516/2008/04/025}{\emph{JCAP} {\bf 04}
  (2008) 025}, [\href{http://arxiv.org/abs/0802.0395}{{\tt 0802.0395}}].

\bibitem{Fujita:2013cna}
T.~Fujita, M.~Kawasaki, Y.~Tada and T.~Takesako, \emph{{A new algorithm for
  calculating the curvature perturbations in stochastic inflation}},
  \href{http://dx.doi.org/10.1088/1475-7516/2013/12/036}{\emph{JCAP} {\bf 12}
  (2013) 036}, [\href{http://arxiv.org/abs/1308.4754}{{\tt 1308.4754}}].

\bibitem{Pattison:2017mbe}
C.~Pattison, V.~Vennin, H.~Assadullahi and D.~Wands, \emph{{Quantum diffusion
  during inflation and primordial black holes}},
  \href{http://dx.doi.org/10.1088/1475-7516/2017/10/046}{\emph{JCAP} {\bf 10}
  (2017) 046}, [\href{http://arxiv.org/abs/1707.00537}{{\tt 1707.00537}}].

\bibitem{Ezquiaga:2019ftu}
J.~M. Ezquiaga, J.~Garc\'\i{}a-Bellido and V.~Vennin, \emph{{The exponential
  tail of inflationary fluctuations: consequences for primordial black holes}},
  \href{http://dx.doi.org/10.1088/1475-7516/2020/03/029}{\emph{JCAP} {\bf 03}
  (2020) 029}, [\href{http://arxiv.org/abs/1912.05399}{{\tt 1912.05399}}].

\bibitem{Grain:2017dqa}
J.~Grain and V.~Vennin, \emph{{Stochastic inflation in phase space: Is slow
  roll a stochastic attractor?}},
  \href{http://dx.doi.org/10.1088/1475-7516/2017/05/045}{\emph{JCAP} {\bf 05}
  (2017) 045}, [\href{http://arxiv.org/abs/1703.00447}{{\tt 1703.00447}}].

\bibitem{Pattison:2019hef}
C.~Pattison, V.~Vennin, H.~Assadullahi and D.~Wands, \emph{{Stochastic
  inflation beyond slow roll}},
  \href{http://dx.doi.org/10.1088/1475-7516/2019/07/031}{\emph{JCAP} {\bf 07}
  (2019) 031}, [\href{http://arxiv.org/abs/1905.06300}{{\tt 1905.06300}}].

\bibitem{Kawasaki:2015ppx}
M.~Kawasaki and Y.~Tada, \emph{{Can massive primordial black holes be produced
  in mild waterfall hybrid inflation?}},
  \href{http://dx.doi.org/10.1088/1475-7516/2016/08/041}{\emph{JCAP} {\bf 08}
  (2016) 041}, [\href{http://arxiv.org/abs/1512.03515}{{\tt 1512.03515}}].

\bibitem{Clesse:2015wea}
S.~Clesse and J.~Garc\'\i{}a-Bellido, \emph{{Massive Primordial Black Holes
  from Hybrid Inflation as Dark Matter and the seeds of Galaxies}},
  \href{http://dx.doi.org/10.1103/PhysRevD.92.023524}{\emph{Phys. Rev. D} {\bf
  92} (2015) 023524}, [\href{http://arxiv.org/abs/1501.07565}{{\tt
  1501.07565}}].

\bibitem{Vennin:2020kng}
V.~Vennin, \emph{{Stochastic inflation and primordial black holes}},
  {\emph{Habilitation thesis} (2020) },
  [\href{http://arxiv.org/abs/2009.08715}{{\tt 2009.08715}}].

\bibitem{Pinol:2020cdp}
L.~Pinol, S.~Renaux-Petel and Y.~Tada, \emph{{A manifestly covariant theory of
  multifield stochastic inflation in phase space}},
  \href{http://arxiv.org/abs/2008.07497}{{\tt 2008.07497}}.

\bibitem{Assadullahi:2016gkk}
H.~Assadullahi, H.~Firouzjahi, M.~Noorbala, V.~Vennin and D.~Wands,
  \emph{{Multiple Fields in Stochastic Inflation}},
  \href{http://dx.doi.org/10.1088/1475-7516/2016/06/043}{\emph{JCAP} {\bf 06}
  (2016) 043}, [\href{http://arxiv.org/abs/1604.04502}{{\tt 1604.04502}}].

\bibitem{Mukhanov:1985rz}
V.~F. Mukhanov, \emph{{Gravitational Instability of the Universe Filled with a
  Scalar Field}}, {\emph{JETP Lett.} {\bf 41} (1985) 493--496}.

\bibitem{Mukhanov:1988jd}
V.~F. Mukhanov, \emph{{Quantum Theory of Gauge Invariant Cosmological
  Perturbations}}, {\emph{Sov. Phys. JETP} {\bf 67} (1988) 1297--1302}.

\bibitem{Tada:2016pmk}
Y.~Tada and V.~Vennin, \emph{{Squeezed bispectrum in the $\delta N$ formalism:
  local observer effect in field space}},
  \href{http://dx.doi.org/10.1088/1475-7516/2017/02/021}{\emph{JCAP} {\bf 02}
  (2017) 021}, [\href{http://arxiv.org/abs/1609.08876}{{\tt 1609.08876}}].

\bibitem{Garcia-Bellido:2017mdw}
J.~Garcia-Bellido and E.~Ruiz~Morales, \emph{{Primordial black holes from
  single field models of inflation}},
  \href{http://dx.doi.org/10.1016/j.dark.2017.09.007}{\emph{Phys. Dark Univ.}
  {\bf 18} (2017) 47--54}, [\href{http://arxiv.org/abs/1702.03901}{{\tt
  1702.03901}}].

\bibitem{Germani:2017bcs}
C.~Germani and T.~Prokopec, \emph{{On primordial black holes from an inflection
  point}}, \href{http://dx.doi.org/10.1016/j.dark.2017.09.001}{\emph{Phys. Dark
  Univ.} {\bf 18} (2017) 6--10}, [\href{http://arxiv.org/abs/1706.04226}{{\tt
  1706.04226}}].

\bibitem{2011ConPh..52..497T}
I.~{Thompson}, \emph{{NIST Handbook of Mathematical Functions, edited by Frank
  W.J. Olver, Daniel W. Lozier, Ronald F. Boisvert, Charles W. Clark}},
  \href{http://dx.doi.org/10.1080/00107514.2011.582161}{\emph{Contemporary
  Physics} {\bf 52} (Sept., 2011) 497--498}.

\bibitem{Martin:2013tda}
J.~Martin, C.~Ringeval and V.~Vennin, \emph{{Encyclop\ae{}dia Inflationaris}},
  \href{http://dx.doi.org/10.1016/j.dark.2014.01.003}{\emph{Phys. Dark Univ.}
  {\bf 5-6} (2014) 75--235}, [\href{http://arxiv.org/abs/1303.3787}{{\tt
  1303.3787}}].

\bibitem{Akrami:2018odb}
{\scshape Planck} collaboration, Y.~Akrami et~al., \emph{{Planck 2018 results.
  X. Constraints on inflation}},
  \href{http://dx.doi.org/10.1051/0004-6361/201833887}{\emph{Astron.
  Astrophys.} {\bf 641} (2020) A10},
  [\href{http://arxiv.org/abs/1807.06211}{{\tt 1807.06211}}].

\bibitem{Aghanim:2018eyx}
{\scshape Planck} collaboration, N.~Aghanim et~al., \emph{{Planck 2018 results.
  VI. Cosmological parameters}},
  \href{http://dx.doi.org/10.1051/0004-6361/201833910}{\emph{Astron.
  Astrophys.} {\bf 641} (2020) A6},
  [\href{http://arxiv.org/abs/1807.06209}{{\tt 1807.06209}}].

\bibitem{Martin:2006rs}
J.~Martin and C.~Ringeval, \emph{{Inflation after WMAP3: Confronting the
  Slow-Roll and Exact Power Spectra to CMB Data}},
  \href{http://dx.doi.org/10.1088/1475-7516/2006/08/009}{\emph{JCAP} {\bf 08}
  (2006) 009}, [\href{http://arxiv.org/abs/astro-ph/0605367}{{\tt
  astro-ph/0605367}}].

\bibitem{Martin:2014nya}
J.~Martin, C.~Ringeval and V.~Vennin, \emph{{Observing Inflationary
  Reheating}},
  \href{http://dx.doi.org/10.1103/PhysRevLett.114.081303}{\emph{Phys. Rev.
  Lett.} {\bf 114} (2015) 081303}, [\href{http://arxiv.org/abs/1410.7958}{{\tt
  1410.7958}}].

\bibitem{Papanikolaou:2020qtd}
T.~Papanikolaou, V.~Vennin and D.~Langlois, \emph{{Gravitational waves from a
  universe filled with primordial black holes}},
  \href{http://arxiv.org/abs/2010.11573}{{\tt 2010.11573}}.

\bibitem{MoradinezhadDizgah:2019wjf}
A.~Moradinezhad~Dizgah, G.~Franciolini and A.~Riotto, \emph{{Primordial Black
  Holes from Broad Spectra: Abundance and Clustering}},
  \href{http://dx.doi.org/10.1088/1475-7516/2019/11/001}{\emph{JCAP} {\bf 11}
  (2019) 001}, [\href{http://arxiv.org/abs/1906.08978}{{\tt 1906.08978}}].

\bibitem{Ooguri:2006in}
H.~Ooguri and C.~Vafa, \emph{{On the Geometry of the String Landscape and the
  Swampland}},
  \href{http://dx.doi.org/10.1016/j.nuclphysb.2006.10.033}{\emph{Nucl. Phys. B}
  {\bf 766} (2007) 21--33}, [\href{http://arxiv.org/abs/hep-th/0605264}{{\tt
  hep-th/0605264}}].

\bibitem{Obied:2018sgi}
G.~Obied, H.~Ooguri, L.~Spodyneiko and C.~Vafa, \emph{{De Sitter Space and the
  Swampland}},  \href{http://arxiv.org/abs/1806.08362}{{\tt 1806.08362}}.

\bibitem{Garg:2018reu}
S.~K. Garg and C.~Krishnan, \emph{{Bounds on Slow Roll and the de Sitter
  Swampland}}, \href{http://dx.doi.org/10.1007/JHEP11(2019)075}{\emph{JHEP}
  {\bf 11} (2019) 075}, [\href{http://arxiv.org/abs/1807.05193}{{\tt
  1807.05193}}].

\bibitem{Ooguri:2018wrx}
H.~Ooguri, E.~Palti, G.~Shiu and C.~Vafa, \emph{{Distance and de Sitter
  Conjectures on the Swampland}},
  \href{http://dx.doi.org/10.1016/j.physletb.2018.11.018}{\emph{Phys. Lett. B}
  {\bf 788} (2019) 180--184}, [\href{http://arxiv.org/abs/1810.05506}{{\tt
  1810.05506}}].

\bibitem{Tada:2019amh}
Y.~Tada and S.~Yokoyama, \emph{{Primordial black hole tower: Dark matter,
  earth-mass, and LIGO black holes}},
  \href{http://dx.doi.org/10.1103/PhysRevD.100.023537}{\emph{Phys. Rev. D} {\bf
  100} (2019) 023537}, [\href{http://arxiv.org/abs/1904.10298}{{\tt
  1904.10298}}].

\bibitem{risken1989fpe}
H.~Risken and H.~Haken, \emph{{The Fokker-Planck Equation: Methods of Solution
  and Applications Second Edition}}.
\newblock Springer, 1989.

\end{thebibliography}\endgroup
\end{document}